\documentclass[12pt]{spieman}  
\usepackage{amsmath,amsfonts,amssymb}
\usepackage{graphicx}
\graphicspath{{f/}}
\usepackage{setspace}

\usepackage{array} 
\usepackage{hyperref}
\usepackage{xurl}

\title{Large-Scale Deployment and Analytical Implications of Structured Quality Control
in Diffusion Magnetic Resonance Imaging}

\author[a]{Michael E. Kim}
\author[b]{Chenyu Gao}
\author[a,b]{Karthik Ramadass}
\author[b]{Gaurav Rudravaram}
\author[b]{Elyssa M. McMaster}
\author[b]{Adam M. Saunders}
\author[e]{Yisu Yang}
\author[b]{Elias Levy}
\author[c]{Praitayini Kanakaraj}
\author[d]{Nancy R. Newlin}
\author[f]{Zhiyuan Li}
\author[g]{Nazirah Mohd Khairi}
\author[h]{Blake E. Dewey}
\author[o]{The HABS-HD Study Team}
\author[p]{Alzheimer’s Disease Neuroimaging Initiative}
\author[i]{Kurt G. Schilling}
\author[j,k]{Derek Archer}
\author[e,l]{Timothy J. Hohman}
\author[a,b,l,m,n]{Bennett A. Landman}
\author[b,*]{Yihao Liu}

\affil[a]{Vanderbilt University, Department of Computer Science, Nashville, TN, USA}
\affil[b]{Vanderbilt University, Department of Electrical and Computer Engineering, Nashville, TN, USA}
\affil[c]{Bioscope.ai, Boston, MA}
\affil[d]{Digital Informatics \& Technology Solutions, Memorial Sloan Kettering Cancer Center, New York, NY}
\affil[e]{Vanderbilt University Medical Center, Vanderbilt Memory and Alzheimer’s Center, Nashville, TN, USA}
\affil[f]{Park University, Department of Computer Science, Parkville, MO, USA}
\affil[g]{Department of Biomedical Data Science, Meharry Medical College, Nashville, TN, USA}
\affil[h]{Department of Neurology, Johns Hopkins University School of Medicine, Baltimore, MD, USA}
\affil[i]{Vanderbilt University Medical Center, Department of Radiology and Radiological Sciences, Nashville, TN, USA}
\affil[j]{Vanderbilt Health, Vanderbilt Memory and Alzheimer’s Center, Nashville, TN, USA}
\affil[k]{Vanderbilt Health, Vanderbilt Genetics Institute, Nashville, TN, USA}
\affil[l]{Vanderbilt University Medical Center, Vanderbilt Genetics Institute, Nashville, TN, USA}
\affil[m]{Vanderbilt University, Department of Biomedical Engineering, Nashville, TN, USA}
\affil[n]{Vanderbilt University Institute of Imaging Science, Nashville, TN, USA}
\affil[o]{HABS-HD MPIs: Sid E O’Bryant, Kristine Yaffe, Arthur Toga, Robert Rissman,
\& Leigh Johnson; and the HABS-HD Investigators: Meredith Braskie, Kevin King, James R Hall, Melissa Petersen,
Raymond Palmer, Robert Barber, Yonggang Shi, Fan Zhang, Rajesh Nandy, Roderick McColl, David Mason,
Bradley Christian, Nicole Phillips, Stephanie Large, Joe Lee, Badri Vardarajan, Monica Rivera Mindt,
Amrita Cheema, Lisa Barnes, Mark Mapstone, Annie Cohen, Amy Kind, Ozioma Okonkwo,
Raul Vintimilla, Zhengyang Zhou, Michael Donohue, Rema Raman, Matthew Borzage, Michelle Mielke, Beau Ances,
Ganesh Babulal, Jorge Llibre-Guerra, Carl Hill and Rocky Vig.}
\affil[p]{Data used in preparation of this article were obtained from the Alzheimer’s Disease
Neuroimaging Initiative (ADNI) database (adni.loni.usc.edu). As such, the investigators
within the ADNI contributed to the design and implementation of ADNI and/or provided
data but did not participate in analysis or writing of this report.
A complete listing of ADNI investigators can be found at: \url{http://adni.loni.usc.edu/wp-content/uploads/how_to_apply/ADNI_Acknowledgement_List.pdf}}

\begin{document}
\maketitle

\begin{abstract}
\textbf{Purpose}: Diffusion MRI~(dMRI) provides a diverse set of quantitative measures and derived datatypes to assess
white matter microstructure and macrostructure. Coupled with the increasing size of imaging studies using dMRI,
the number of downstream outputs requiring quality control~(QC) will continue to grow.
Previous work has shown that failure modes which are often not evident from aggregate metrics or summary
statistics can be identified through structured visual inspection.
This work aims to better understand common failure modes and the expected characteristics of valid dMRI processing outputs to ensure the validity and
interpretability of quantitative findings.
\textbf{Approach}:
We deployed a structured QC framework to assess 18{,}328 dMRI scans across nine datasets,
visually evaluating the outputs of seven processing pipelines representative of conventional dMRI analyses.
\textbf{Results}:
Downstream outputs that pass visual QC may still rely on failed upstream dependencies;
such failures may only be visually detectable through systematic inspection of the full pipeline hierarchy.
Additionally, appropriate QC granularity is algorithm-specific, as the spatial structure of each algorithm’s outputs determines
whether failures warrant selective or global exclusion.
\textbf{Conclusion}:
This work demonstrates the feasibility and analytical value of large-scale, structured QC for dMRI processing pipelines.
Our results highlight the need for systematic QC spanning the full processing hierarchy to ensure the validity and interpretability of quantitative findings.
\end{abstract}


{\noindent \footnotesize\textbf{*}Yihao Liu, \linkable{yihao.liu@vanderbilt.edu}}

\begin{spacing}{2}

\section{Introduction}
Diffusion weighted magnetic resonance imaging~(dMRI) provides a powerful noninvasive window into the structural organization of the human brain.
By characterizing the directional self-diffusion of water molecules within tissue,
dMRI enables the investigation of white matter microstructure and large-scale connectivity patterns~\cite{jones2011diffusion}.
Over the past two decades, advances in acquisition protocols, modeling approaches,
and computational pipelines have made it possible to derive increasingly rich
quantitative representations of white matter architecture~\cite{novikov2019quantifying}.
As a result, modern neuroimaging studies frequently integrate dMRI with structural imaging to generate a wide variety of
downstream derivative data products used in neuroscience and clinical research~\cite{kamagata2024advancements,inglese2010diffusion}.
These derivatives span multiple classes of measurements, including tensor-derived scalar maps such as
fractional anisotropy~(FA) and mean diffusivity~(MD)~\cite{alexander2007diffusion},
atlas-based white matter segmentations~\cite{hansen2021pandora,yang2022learning,oishi2008human,oishi2009atlas,mori2008stereotaxic},
tractography streamline-based bundle segmentations~\cite{behrens2007probabilistic,watanabe2018detection,wasserthal2018tractseg},
structural connectivity matrices and graph measures derived from these matrices~\cite{sporns2018graph,yeh2021mapping},
and microstructural models that characterize different properties of white matter integrity and
cellular architecture~\cite{alexander2019imaging,novikov2019quantifying}. 

These measurements have shown great promise in the investigation of diseases and disorders,
and as clinical and research data become more available, researchers have transitioned
to incorporate these dMRI measurements at unprecedented scales.
Several works have emerged aggregating multiple datasets for mega-analyses of white matter across the human lifespan
to create normative reference frameworks and investigate notable
periods of population-level change~\cite{kim2025white,zhu2025lifespan,schilling2023white}.
Beyond population modeling, these large-scale studies allow researchers to robustly detect small
biological effects unobservable within smaller sample sizes~\cite{feng2022effect}
and reliably estimate rapid development and growth during developmental ages~\cite{dubois2021mri}.
Several consortia have formed to address these mega-studies of imaging data,
such as the Enhancing NeuroImaging Genetics through Meta Analysis~(ENIGMA) consortium~\cite{thompson2020enigma},
the HEALthy Brain and Child Development~(HBCD) Study~\cite{nelson2024introduction},
and the Alzheimer’s Disease Sequencing Project – Phenotype Harmonization Consortium~(ADSP-PHC)~\cite{hohman2023adsp}. 

With the increased use of these measurements, effective quality control~(QC) at scale becomes paramount.
Currently, there are many approaches to determine quality of diffusion MRI data used for research studies.
Many methods rely on automated approaches for outlier detection
to reduce the manual effort on QC~\cite{liu2010quality,samani2020qc,sanchez2023fetmrqc,gedamu2008automated,esteban2017mriqc}.
However, other researchers suggest that manual QC remains the gold standard,
as relying on statistical or deep learning QC can lead to poor quality
data leaking into research studies~\cite{cieslak2025diffusion,kim2025scalable}.
While the debate between manual and automated methods focuses on how data quality is evaluated,
another critical consideration is when and where these checks occur within the broader processing pipeline.
In practice, these outputs are produced through multi-stage computational workflows that involve preprocessing,
tensor fitting, atlas registration, tractography, and multi-compartment modeling.
Each stage introduces assumptions and potential algorithmic failure modes,
including gradient metadata errors, motion-correction artifacts, registration misalignment,
tractography failures, and model-fitting instability.
Since downstream analyses depend on the successful execution of earlier pipeline stages,
undetected processing errors can propagate and
influence derived quantitative measures~\cite{power2012spurious,backhausen2016quality,ducharme2016trajectories}.
Several prior works focus on QC at only a single point in an interdependent pipeline,
either for the raw data~\cite{esteban2017mriqc}, the outputs of preprocessing steps~\cite{esteban2019fmriprep},
or the final, post-processed outputs of image processing pipelines~\cite{HE2021109099}.
Many approaches for dMRI focus on QC of the preprocessed data, with some frameworks
integrating QC as part of the pipeline itself~\cite{prequal,dwiqc}. 

Despite the importance of these dMRI-derived outputs and the need for rigorous quality control,
there is a lack of empirical studies that characterize how upstream processing failures
propagate to downstream outputs.
Such information could reveal the extent to which downstream derivatives depend on the integrity of earlier pipeline
and reveal processing pipelines that are most susceptible
to error propagation, thereby serving as QC bottlenecks.
Thus, in this work we describe the large-scale deployment of a structured QC
framework designed to evaluate diffusion-derived outputs across multiple stages of contemporary neuroimaging workflows.
The framework was applied to derivatives from commonly used neuroimaging pipelines spanning preprocessing, tissue segmentation, tensor modeling, atlas registration, tractography, and microstructural modeling.
The QC approach emphasizes standardized visualization outputs, rapid inspection of algorithm-specific diagnostics, and structured documentation of QC outcomes to enable scalable evaluation across large datasets.

We deployed this framework across nine neuroimaging datasets encompassing 18{,}328 scans from 10{,}806 subjects,
evaluating outputs from seven commonly used processing pipelines.
For each pipeline stage, we describe the expected visual characteristics of valid outputs and
document the algorithm-specific failure modes encountered in practice.
Our results demonstrate that many characteristic failure modes are identifiable through systematic visual inspection.
Critically, a non-trivial proportion of outputs that pass visual QC rely on upstream dependencies that have themselves failed,
highlighting that QC must span the full pipeline hierarchy rather than individual stages in isolation.
Finally, we show that the appropriate granularity of QC is algorithm-specific:
for spatially structured outputs, per-region evaluation enables selective exclusion of affected areas and retain usable data. 

\section{Quality Control Approach and Common Failure Modes}
\label{sec:method}
Our QC framework builds upon a previously published web-based interface designed for scalable and standardized visual assessment of neuroimaging data~\cite{kim2025scalable}.
Within this framework, users are presented precomputed visualizations and diagnostic outputs in a consistent format,
which enables rapid and uniform evaluation without requiring interaction with multiple specialized software tools.
Assessments are performed with predefined criteria to promote consistency across raters.
This section describes the QC procedures applied to each processing component,
focusing on the construction of standardized visualization outputs, the expected anatomical,
algorithm-specific patterns to assess data integrity, and the common failure modes.
We provide representative visualizations of these expected patterns and typical failure cases to support interpretation.

\subsection{PreQual}

The first step in diffusion-weighted imaging (DWI) data analysis is preprocessing
to correct image artifacts and ensure consistency in gradient metadata information~\cite{tax2022s}.
When performing QC on the PreQual preprocessing pipeline~\cite{prequal} for DWI data, there are several outputs that we focus on to determine if there is an issue with the data,
since PreQual itself is an amalgamation of several different algorithms and tools.
As PreQual steps are partly focused on ensuring the data are properly formatted,
the QC mainly revolves around the original data quality and associated metadata needed to run the pipeline.
An example of PreQual QC page is shown in Supplementary Figs.~\ref{fig:prequal_pdf}.

First, we look at the outputs of the EDDY tool from FSL~\cite{jenkinson2012fsl} and intensity across volumes.
For any 3D volume in a 4D DWI image, the intensity is related to the biological environment of the tissue and the gradient applied for diffusion-weighting.
The signal intensity $S$ at any voxel in the image can be described as:
\begin{equation}
\label{eq:diffusion}
S = S_{0}\,e^{-b\, \mathbf{D}} \, ,
\end{equation}
where $S_{0}$ is the signal intensity without diffusion-weighting, $b$ is the b-value that
represents the strength of the signal attenuation,
and $\mathbf{D}$ is the coefficient for the self-diffusion of water molecules.
Thus, the median intensity within a 3D volume is expected to decay exponentially with respect to the b-value.
We expect motion during in vivo human scans, and thus,
plots of estimated translation and rotation of the head over time should be relatively smooth between adjacent volumes.
For the EDDY slice-wise imputation algorithm, slices detected as outliers are replaced with a ``prediction'' by the algorithm;
an output with an unreasonable number of slices replaced, especially in central brain slices, would be a cause for concern in QC.
Finally, we examine the chi-square analysis of the observed signal compared to the signal upon reconstruction from a modeled diffusion tensor,
with the expectation that volumes with a b-value equal to or less than 1500 should be well-approximated by the tensor model.

Next, we look at the tensor visualization and sampled b-vector orientations,
where b-vector is a term that refers to the unit normalized direction vector of the applied gradient pulse for diffusion weighting.
While modeled diffusion tensors are visualized for every voxel,
errors can be easily identified upon visual inspection within two main regions of interest:
the corpus callosum for an axial slice and the corticospinal tract for a coronal slice.
In an axial slice, the corpus callosum tensors should create a ``U'' shape and be oriented left-to-right (Supplementary Fig.~\ref{fig:prequal}),
whereas the corticospinal tract tensors should mostly run superior-to-inferior on a coronal slice view.
Additionally, PreQual performs a permutation check of the b-vectors to determine if there is an order and sign permutation
that provides a more optimal streamline propagation for tractography along the direction of the diffusion tensor orientation.
If the b-vectors are properly oriented, then the optimal orientation will be the original.

When there is no reverse phase-encoding scan available to perform susceptibility-induced distortion correction with TOPUP,
PreQual uses a deep-learning method called Synb0-DisCo~\cite{schilling2019synthesized,schilling2020distortion} to generate a synthetic 3D image from a paired T1w scan that can be supplied to TOPUP to correct the distortions.
For these instances, we also visualize the synthesized image to ensure that it looks anatomically accurate for use in correction.
Finally, we also visualize an FA map created from a tensor using all image volumes,
where we expect the white matter to appear bright compared to the rest of the brain.

Any issues regarding the metadata of a scan, for instance, improperly reported b-values or b-vectors,
are oftentimes common for an entire dataset. To correct these discrepancies,
we advocate for a few preliminary pipeline runs with QC prior to parallelization of preprocessing for the entire dataset.
The most frequently encountered issue with these discrepancies is a flip of the reported b-vectors across the x-axis,
which can be easily verified from the tensor orientation of the corpus callosum and comparison to the
optimal b-vector orientation (Supplementary Fig.~\ref{fig:prequal}).
Another common error is an improper synthetic image from Synb0-DisCo,
where the synthesized image has anatomical inaccuracies or is not a full
brain (Supplementary Figs.~\ref{fig:supp1},~\ref{fig:supp2}).
An indicator of a failed EDDY process is when the estimated motion is hallucinated by the algorithm,
resulting in translation or rotation plots where motion is sporadic and incredibly discontinuous (Supplementary Fig.~\ref{fig:prequal}).
Less frequent examples of errors are improper b-value reporting,
which can be easily seen by comparing median intensity and b-value across volumes,
and high chi-square values across multiple center slices of the brain for all volumes under 1500 b-values.
While the FA map and slice imputation analysis help to provide additional assurance of a failed instance,
issues with these two QC steps most often present with other issues rather than alone.

\subsection{T1-weighted Segmentation}

Downstream DWI processing can require region-of-interest (ROI) parcellation,
which is often performed on higher-resolution T1-weighted MRI data.
For SLANT-TICV~\cite{liu2022generalizing}, UNesT~\cite{yu2023unest}, or MaCRUISE~\cite{huo2016consistent},
which are segmentation algorithms that mostly focus on the gray matter of the brain,
an overlay with slight opacity for the label map can usually indicate success of the algorithm.
The expectation is that the cortex of the brain is properly parcellated,
with all labels in their approximate corresponding positions (Supplementary Fig.~\ref{fig:segmentation}).
While failures are infrequent, the most common errors that do occur are when gray matter labels leak outside
of the brain or into the white matter and vice versa.

\subsection{Tensor Fitting and Atlas Registration}

An alternative approach to ROI segmentation is atlas registration.
For registration of the JHU template~\cite{oishi2010jhu,oishi2009atlas}, we visualize the transformed EVE3 labels in the subject space on top of an FA map
extracted from a tensor fit using volumes with b-values less than 1500~\cite{tournier2019mrtrix3},
which has been masked with the binarized SLANT-TICV/UNesT T1w image segmentation.
We expect the registration to result in the EVE3 labels roughly aligning with the white matter
tracts present on the FA map (Supplementary Fig.~\ref{fig:eve3}).
Typical failure cases can be identified with visual inspection and include the labels or
applied mask appearing off-center from the FA map.

\subsection{Tractseg}

One benefit of DWI data is the ability to reconstruct WM pathways in the brain through tractography~\cite{jeurissen2019diffusion}.
As Tractseg~\cite{wasserthal2018tractseg}, a bundle-specific tractography algorithm,
outputs 72 different white matter bundles that have substantial spatial overlap with each other,
we visualize each bundle in a separate PNG. Visualizing all bundles, or tracts,
simultaneously would make interpretation of quality difficult.
All tracts are expected to have successfully generated enough streamlines to give a ``full'' appearance to the white matter
bundle (Supplementary Fig.~\ref{fig:tractseg}).

Most encountered issues with outputs are a small number of streamlines that make the bundle appear wispy,
or entirely empty bundles due to failures in upstream processing steps.
A complete lack of streamlines will be caught in later processing when calculating the average
tensor scalar metrics within the bundle, as the calculation of the average will be infeasible,
but the QC identifies edge cases with infeasible but nonempty bundles.
Much rarer issues are when streamlines are improperly located or terminate early.

\subsection{Pasternak Single-Shell Freewater Estimation}

DWI microstructure can be modeled beyond the canonical diffusion tensor,
with a common alternative being the bi-tensor estimation of a tissue and free water component.
For visualization of the free water estimation~\cite{pasternak2009free}, we look at the free water-corrected FA maps compared to the uncorrected versions.
We expect the free water correction to increase the relative intensity of the white matter pathways without alteration to their overall structure.
A common failure is when the free water correction results in an FA map that is much noisier than the original or when the model
fit results in an overestimation of the free water-corrected FA values in non-WM regions (Supplementary Fig.~\ref{fig:supp5}).

\subsection{White Matter Brain Ages}

The concept of an imaging “brain age”, a measure for brain health that is based on the expected age from an MRI scan,
has been used for several imaging modalities~\cite{sajedi2019age}. For DWI-specific brain ages,
we use the white matter brain age (BRAID) algorithm~\cite{braid}.
As the output for a white matter brain age is a single scalar value of the brain age,
there is no qualitative visualization of the outputs on an individual level.
However, the FA/MD maps and the T1w scan registered to the MNI template can be inspected as
part of the upstream pipeline (Supplementary Fig.~\ref{fig:supp7}).
For the MNI 152~\cite{fonov2009unbiased,tustison2014large} registration, we visualize the MNI 152 template and FA/MD maps,
but with affine and deformable transformations applied to place them in MNI 152 space.
Both have been masked with the binarized SLANT-TICV/UNesT T1w image segmentation.
We expect minimized macrostructural information for the deformable transformation and
relative alignment for the affine only registration.

\subsection{Connectome Special}

A unique feature set from DWI data is structural brain connectivity,
which estimates the ROI-based connectivity across the brain using whole-brain tractography~\cite{behrens2012human}.
The main output of the Connectome Special~\cite{rubinov2010complex,tournier2019mrtrix3}, i.e.\ the structural connectome matrix, is the focus of the QC for the pipeline outputs.
We expect the connectomes to be diagonally symmetric, with a reasonable number of strong connections between brain regions, visually appearing as a more “full” matrix.
The main diagonal should have the strongest connections for the connectome weighted by the number of streamlines~(NOS).
Additionally, the connectome weighted by the average FA value should have a relatively homogeneous intensity (Supplementary Fig.~\ref{fig:connectome}).
Most issues that occur present when the main diagonal is not visible for the NOS-weighted connectome or if the FA-weighted
connectome has large regions of inhomogeneity.

The seven pipelines above were deployed at scale across the datasets described in Section~\ref{sec:result}.
Additional pipelines including BISCUIT~\cite{lyu2017novel,lyu2018cortical}, Freesurfer~\cite{fischl2012freesurfer}, NODDI~\cite{zhang2012noddi}, and Nextflow Bundle Analysis~\cite{yeatman2012tract,theaud2020tractoflow,rheault2020analyse,garyfallidis2018recognition,di2017nextflow,cousineau2017test,cote2015cleaning}
were also studied on a subset of data and their QC procedures are documented in the Supplemental Material.

\section{Results}
\label{sec:result}
We deployed the structured QC framework across nine neuroimaging datasets:
ADNI~\cite{weber2021worldwide}, HABS-HD~\cite{o2021health},
NACC, FloridaADRC, Indiana~\cite{beekly2007national}, ROSMAPMARS~\cite{l2012minority,bennett2005rush,a2012overview},
SCAN~\cite{scan_dataset}, WASHU~\cite{fernandez2024genetic}, and WRAP~\cite{sager2005middle},
encompassing a total of 18{,}328 scans from 10{,}806 subjects across 17{,}763 sessions.
All raw data passed QC prior to our analysis, isolating the pipeline-specific failures of interest from data quality issues.
Raw data QC consisted of visual checks to confirm that the scan was of the expected modality and was free of major image artifacts.
The 18{,}328 scans were processed through the seven pipelines described in Section~\ref{sec:method},
QC was distributed across a team of trained raters, with each pipeline-generated output reviewed by a single rater.
QC outcomes were then aggregated across all pipelines.

\begin{figure}
    \centering
    \begin{tabular}{cc}
        Pipeline Dependency Chain
        \\
         \includegraphics[width=0.7\linewidth, trim=390pt 150pt 580pt 200pt, clip]{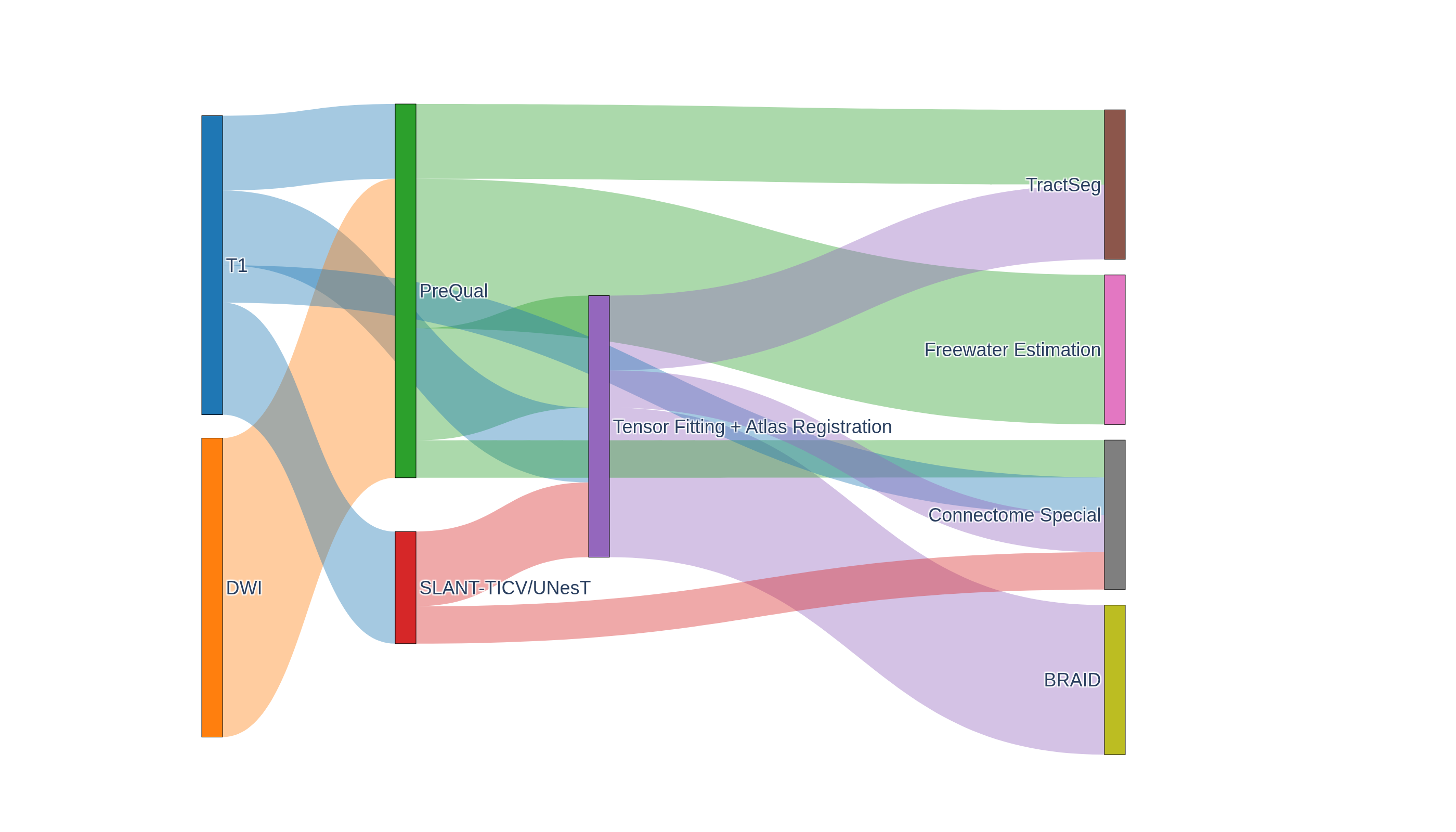}
         \\

             \begin{tabular}{>{\centering\arraybackslash}m{0.18\textwidth}
                    >{\centering\arraybackslash}m{0.18\textwidth}
                    >{\centering\arraybackslash}m{0.18\textwidth}
                    >{\centering\arraybackslash}m{0.18\textwidth}
                    >{\centering\arraybackslash}m{0.18\textwidth}
            }
            PreQual & SLANT-TICV/UNesT & Freewater Estimation & \begin{tabular}{c}
                 Tensor Fitting  \\
                 Atlas Registration 
            \end{tabular}
            & BRAID
            \\
            \includegraphics[width=0.18\textwidth, trim=40pt 40pt 40pt 40pt, clip]{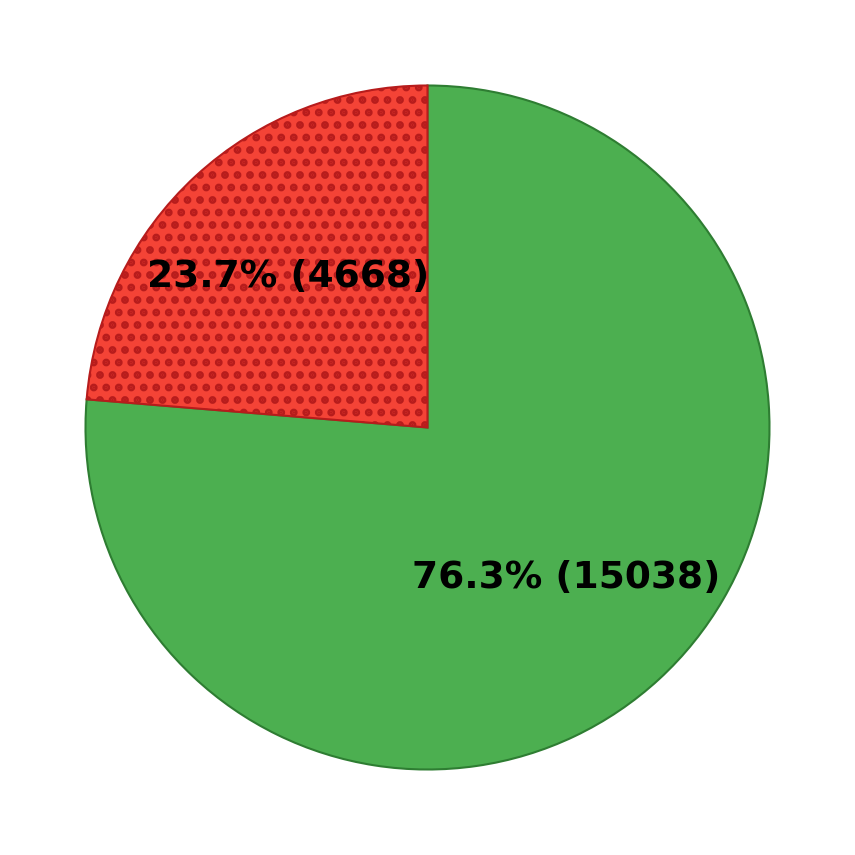}
            &
            \includegraphics[width=0.18\textwidth, trim=40pt 40pt 40pt 40pt, clip]{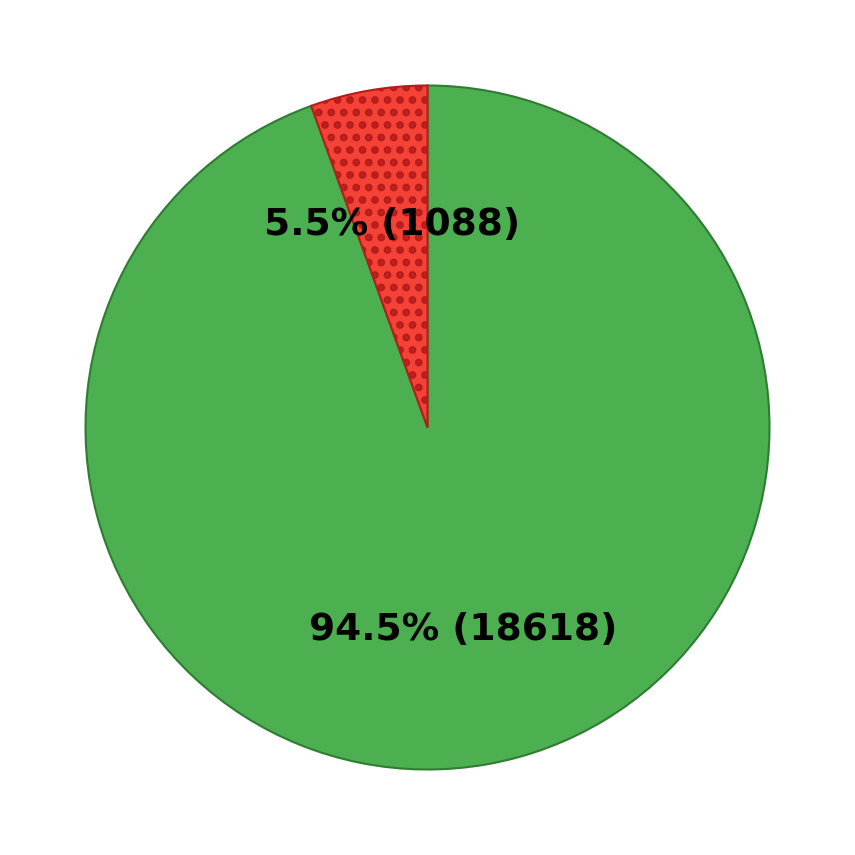}
            &
            \includegraphics[width=0.18\textwidth, trim=40pt 40pt 40pt 40pt, clip]{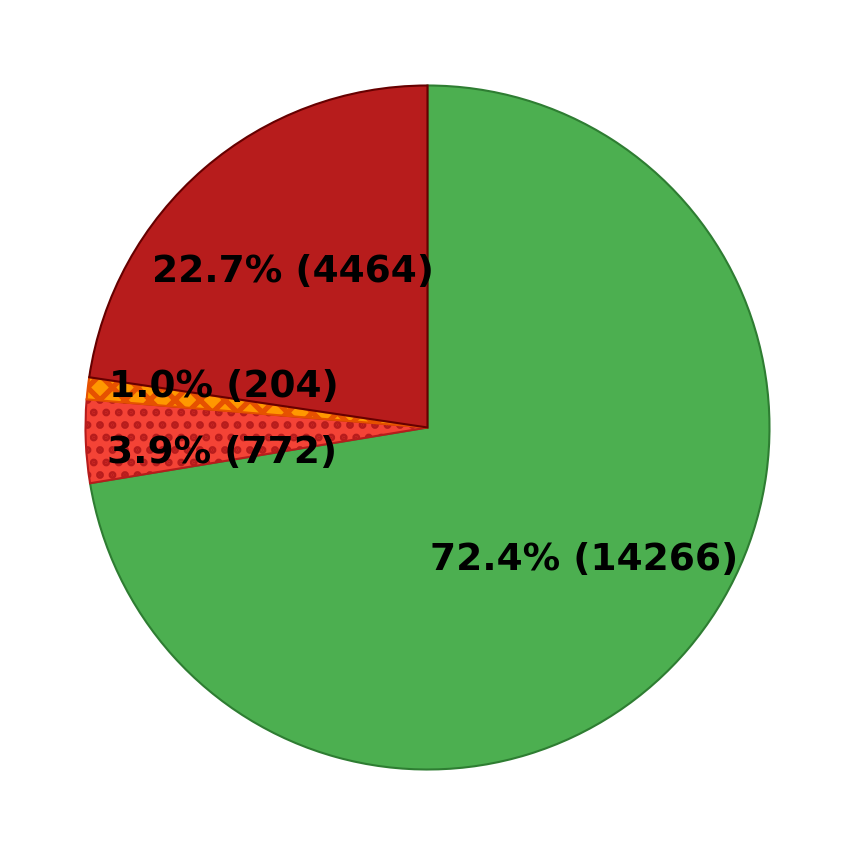}
            &
            \includegraphics[width=0.18\textwidth, trim=40pt 40pt 40pt 40pt, clip]{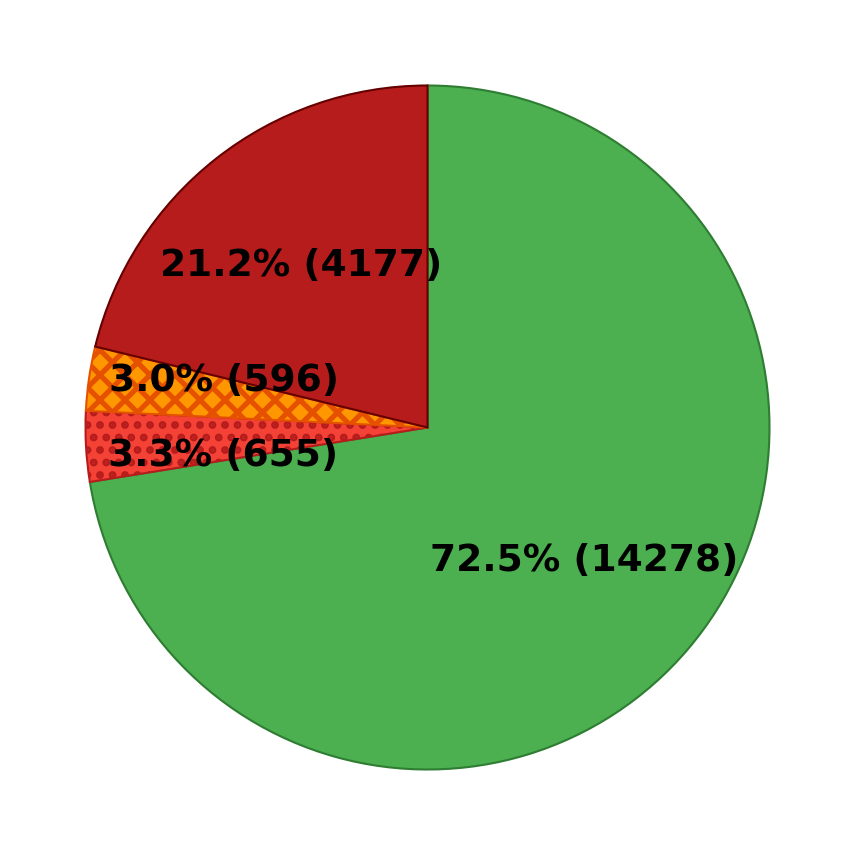}
            &
            \includegraphics[width=0.18\textwidth, trim=40pt 40pt 40pt 40pt, clip]{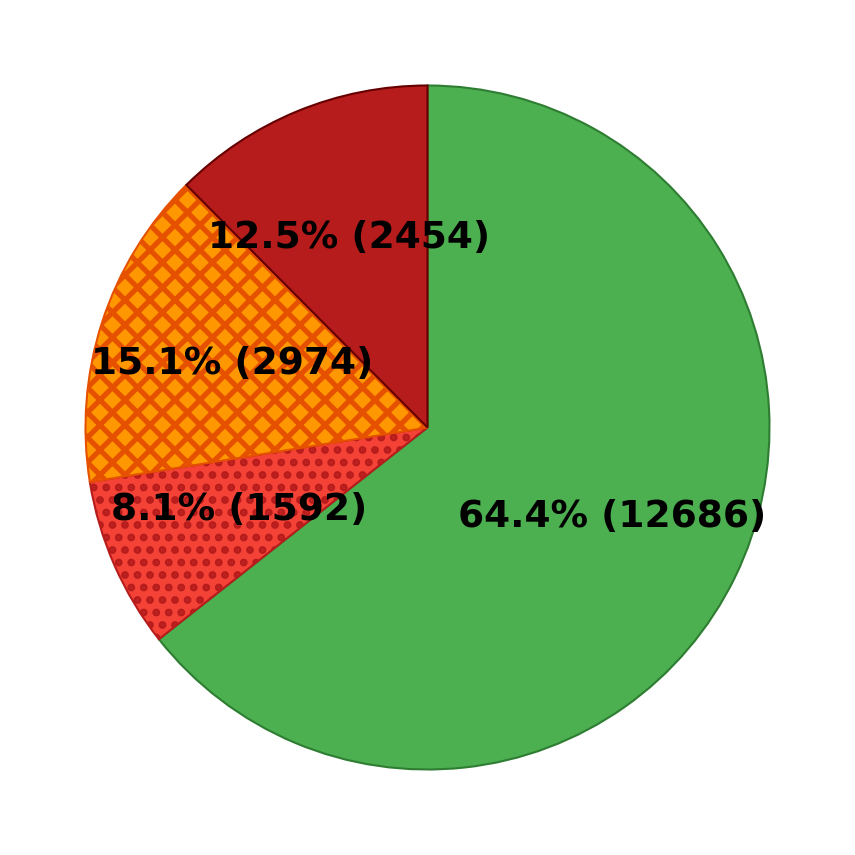}
            \\
            \begin{tabular}{c}
                 Connectome  \\
                 Special 
            \end{tabular}
            &
            \begin{tabular}{c}
                 Tractseg  \\
                 AF Right 
            \end{tabular}
            &
            \begin{tabular}{c}
                 Tractseg  \\
                 ATR Left
            \end{tabular}
            &
            \begin{tabular}{c}
                 Tractseg  \\
                 CC5
            \end{tabular}
            \\
            \includegraphics[width=0.18\textwidth, trim=40pt 40pt 40pt 40pt, clip]{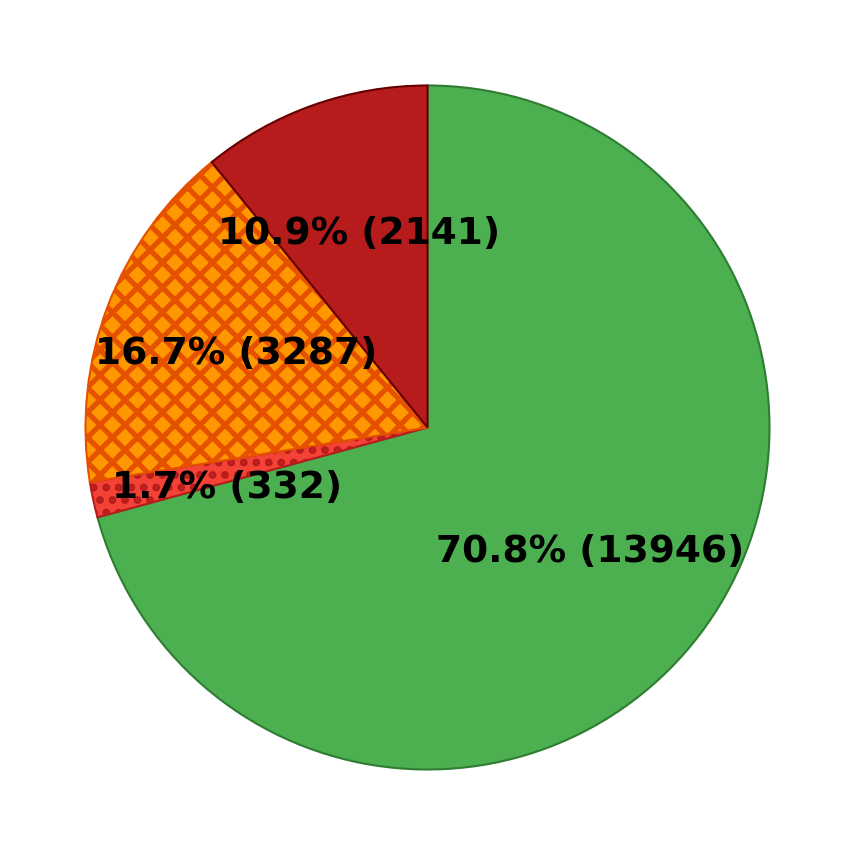}
            &
            \includegraphics[width=0.18\textwidth, trim=40pt 40pt 40pt 40pt, clip]{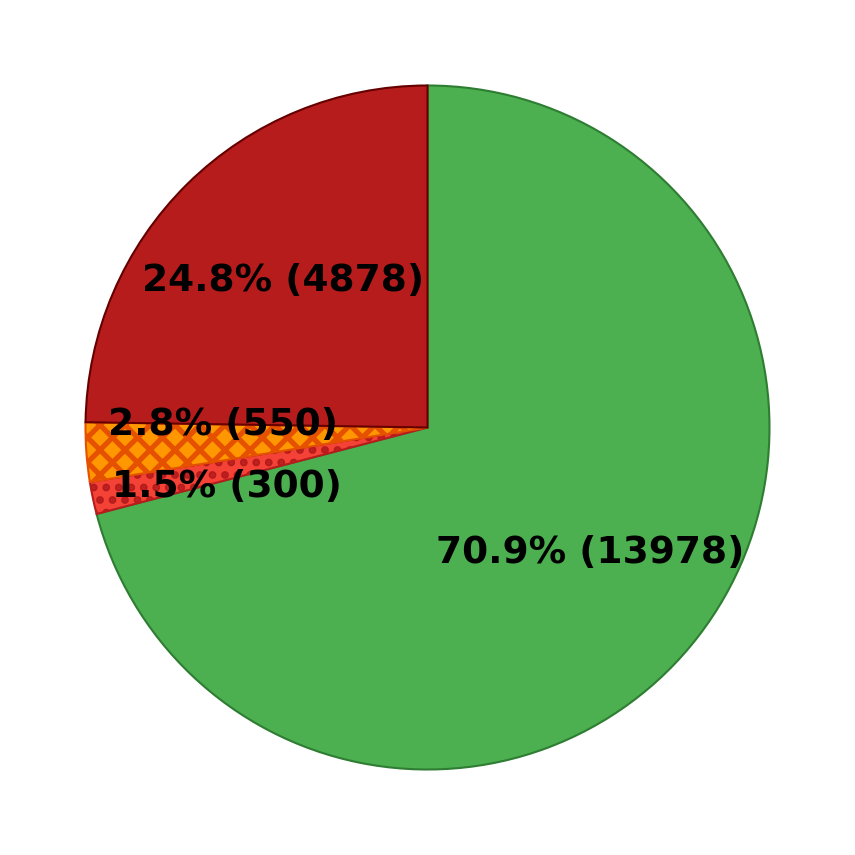}
            &
            \includegraphics[width=0.18\textwidth, trim=40pt 40pt 40pt 40pt, clip]{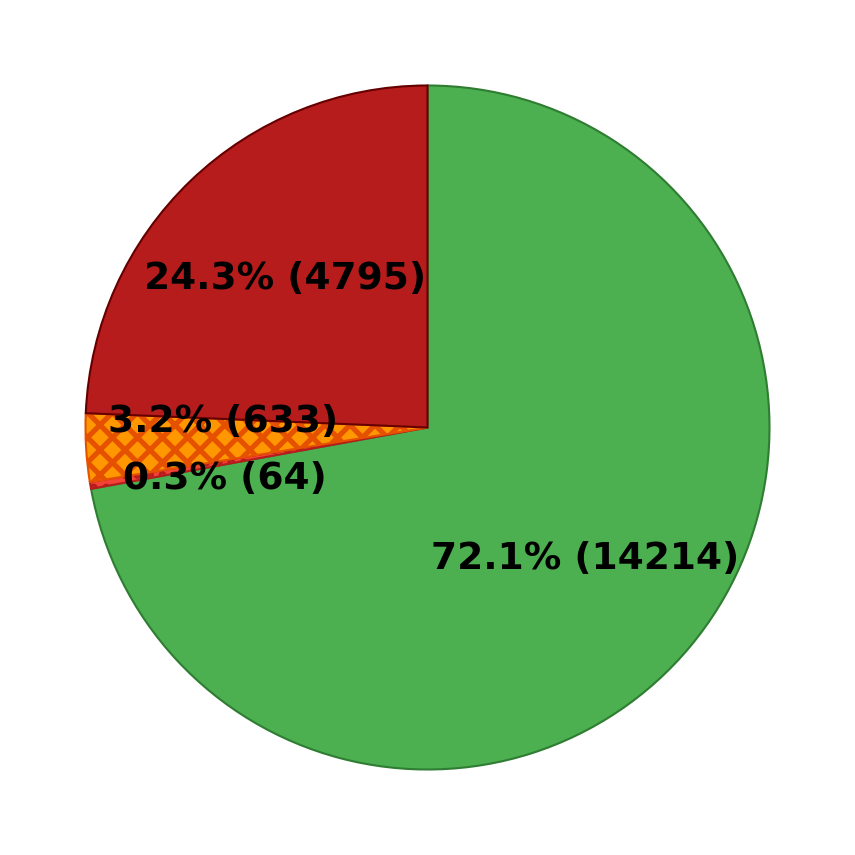}
            &
            \includegraphics[width=0.18\textwidth, trim=40pt 40pt 40pt 40pt, clip]{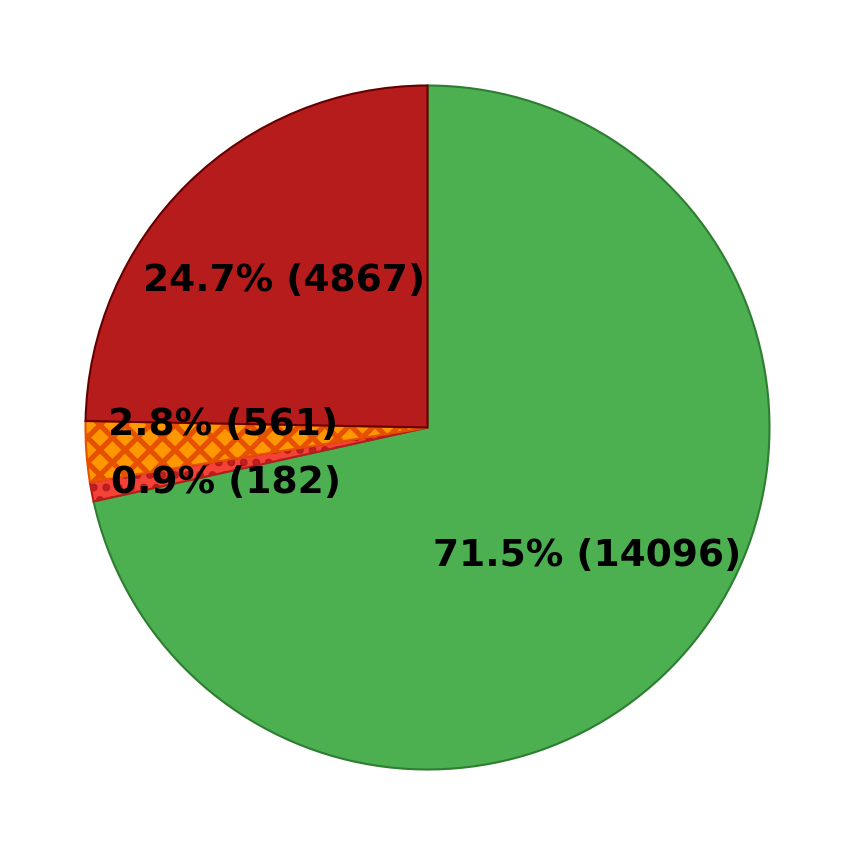}
            &
            \includegraphics[width=0.18\textwidth]{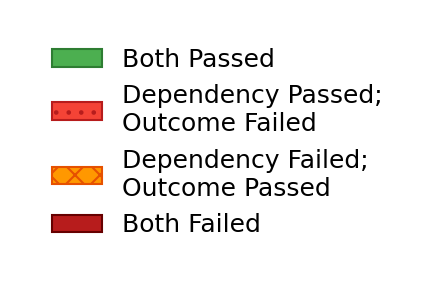}
    
        \end{tabular}
    
    \end{tabular}
    
    \caption{\label{f:pie}
        Processing pipeline hierarchy and QC outcomes for each processing stage.
        Three tracts from Tractseg are shown: right arcuate fasciculus (AF Right), left anterior thalamic radiation (ATR Left),
        and posterior midbody of the corpus callosum (CC5). Each pie chart represents the proportion of scans by QC outcome category.
        Both Passed: both the processing step and all its dependencies passed QC. Dependency Passed; Outcome Failed: all dependencies passed QC,
        but the processing step failed. Dependency Failed; Outcome Passed: the processing step passed QC, but at least one dependency failed.
        Both Failed: both the processing step and one or more dependencies failed.
        Downstream pipelines illustrate the cumulative impact of dependency failures.
    }
\end{figure}

Figure~\ref{f:pie} summarizes the QC outcomes across all processing stages.
Because Tractseg produces 72 individual tract segmentations, three representative bundles are shown: right arcuate fasciculus~(AF~Right), left anterior thalamic radiation~(ATR~Left), and posterior midbody of the corpus callosum~(CC5).
Results are categorized into four mutually exclusive outcome classes: (1)~\textit{Both Passed}: both the processing step and all upstream dependencies passed QC;
(2)~\textit{Dependency Passed; Outcome Failed}: all upstream dependencies passed QC, but the processing step itself failed;
(3)~\textit{Dependency Failed; Outcome Passed}: the processing step passed QC, but at least one upstream dependency failed;
and (4)~\textit{Both Failed}: both the processing step and one or more upstream dependencies failed QC.
We note that instances where the pipeline was unable to be run or unable to be run to completion are considered failures.

\section{Discussion}

\subsection*{Inspecting downstream outputs alone cannot always reveal dependency failures}

\begin{figure}[ht]
    \begin{center}
        \begin{tabular}{c}
        \includegraphics[width=\linewidth]{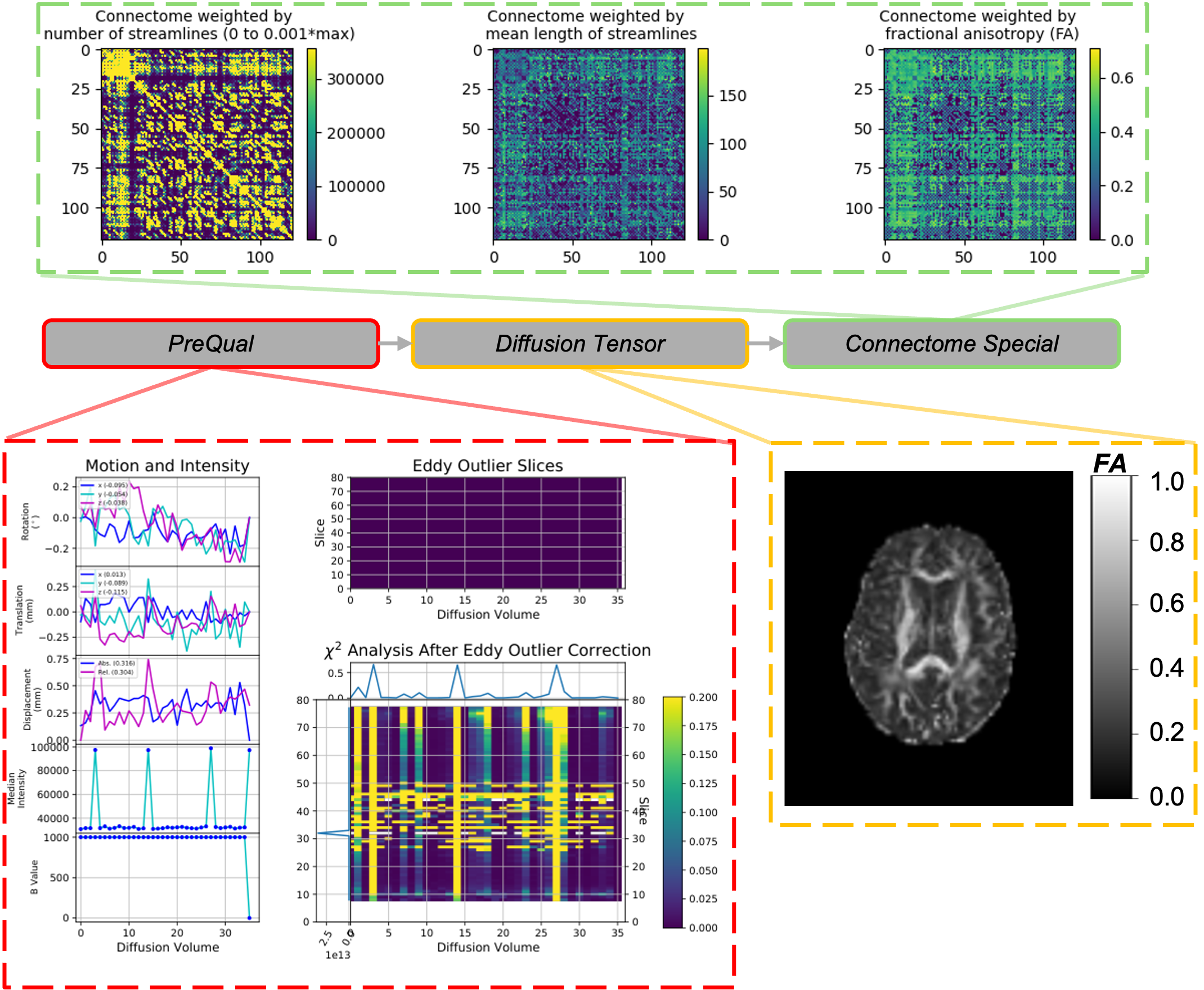}
        \end{tabular}
    \end{center}
    \caption{\label{fig:dependency_1}
        Case example of a Connectome Special output that passed QC, but for which dependencies did not pass QC.
        Notably, the PreQual pipeline outputs indicated problematic slices throughout a majority of volumes for slices 30 to 50,
        which are in the middle of the brain. Furthermore, the downstream tensor fit showcases high FA values in non-white
        matter brain regions, indicating questionable data quality. However, the structural connectomes appear visually reasonable.
    }
\end{figure}

\begin{figure}[ht]
    \begin{center}
        \begin{tabular}{c}
        \includegraphics[width=\linewidth]{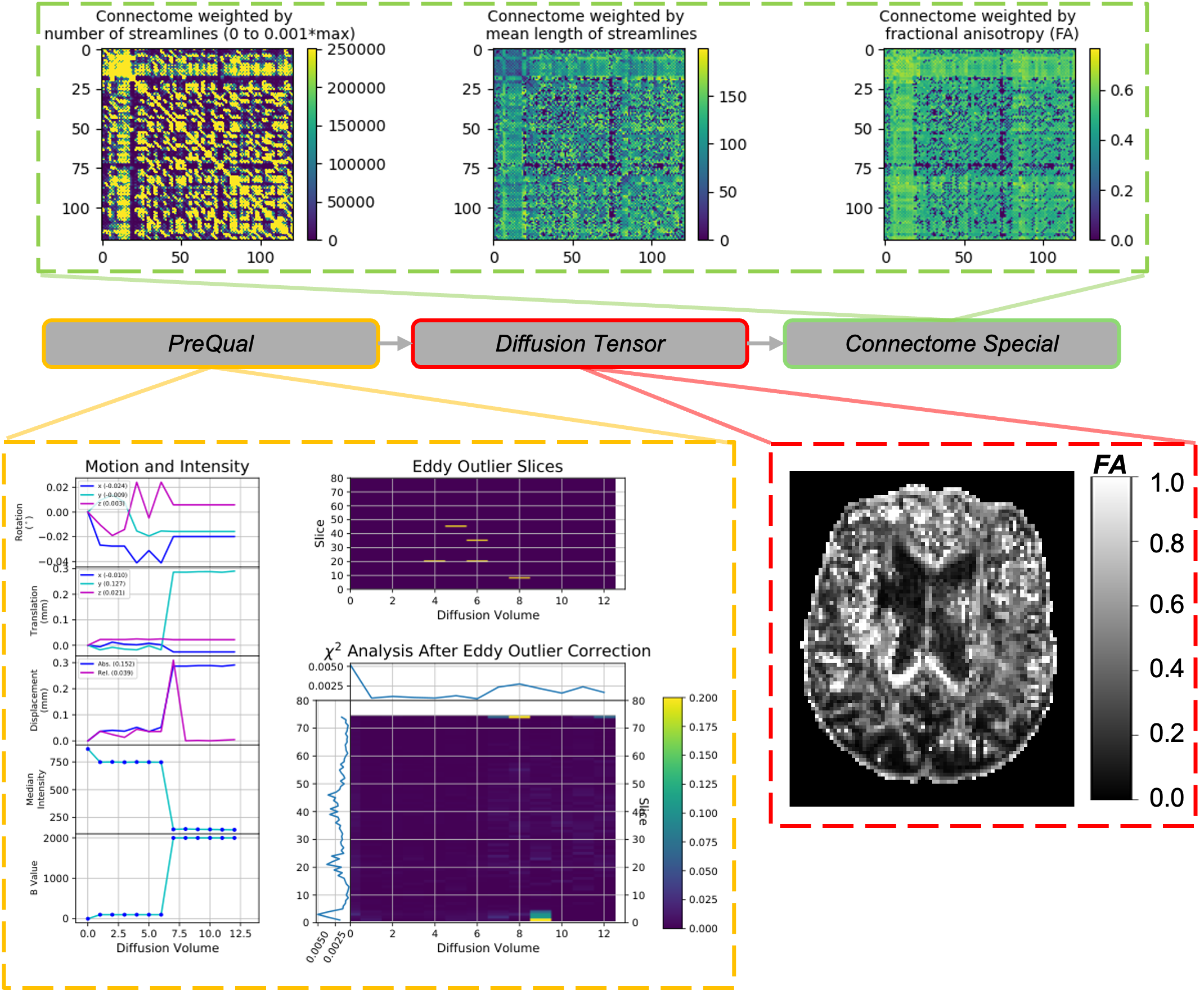}
        \end{tabular}
    \end{center}
    \caption{\label{fig:dependency_2}
    Case example of a Connectome Special output that passed QC, but for which dependencies did not pass QC.
    Notably, the outputs of the PreQual pipeline indicate the scan had reasonable motion and data quality;
    however, the number of volumes in the data (12, not including a single b0 volume) is less than the number
    typically used in literature~\cite{tournier2011diffusion}.
    As only six of these are below b-values of 1500, the resulting tensor fit is also unstable,
    resulting in a poor-quality FA image.
    Despite this, the downstream connectivity matrices appear visually high-quality.
    }
\end{figure}

\begin{figure}[ht]
    \begin{center}
        \begin{tabular}{c}
        \includegraphics[width=\linewidth]{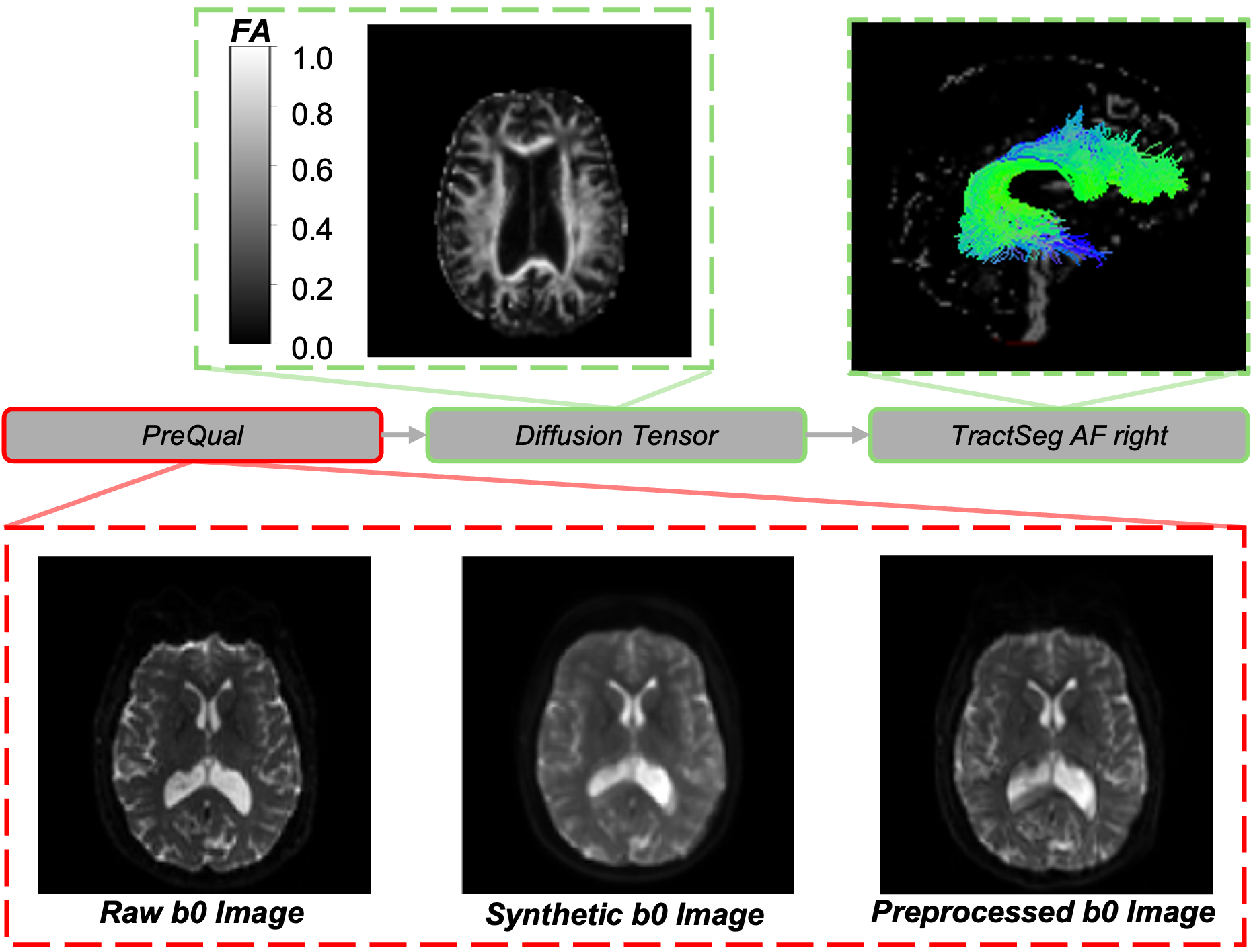}
        \end{tabular}
    \end{center}
    \caption{\label{fig:dependency_3}
    Case example of a Tractseg (AF right) output that passed QC and had a dependency (tensor) pass QC,
    but for which the PreQual dependency did not pass QC. While the FA map and reconstructed tract both look reasonable,
    the preprocessed scan is problematic because of an issue with the synthetic b0 image used for
    susceptibility-induced distortion correction.
    }
\end{figure}

\begin{figure}[ht]
    \begin{center}
        \begin{tabular}{c}
        \includegraphics[width=\linewidth]{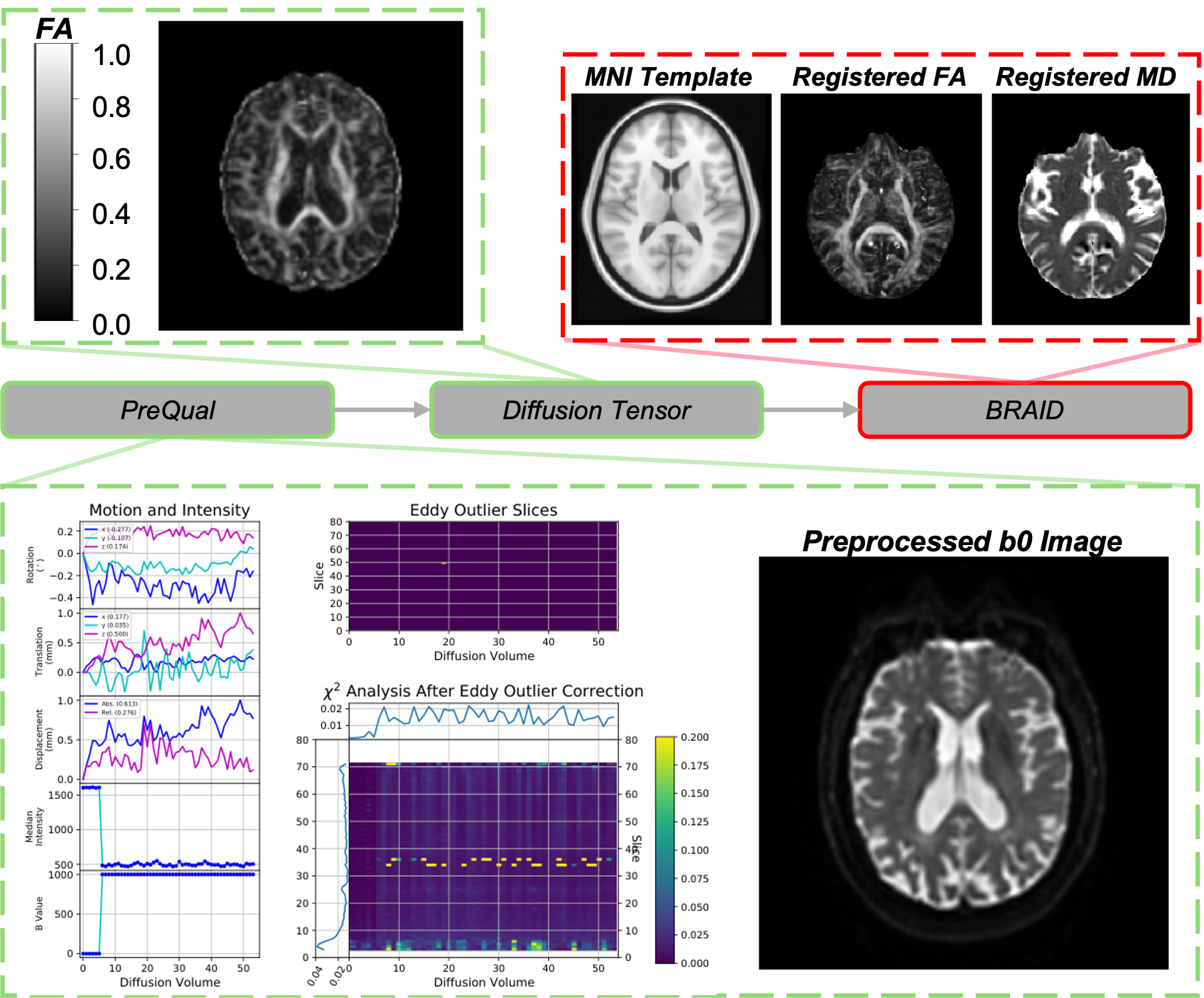}
        \end{tabular}
    \end{center}
    \caption{\label{fig:dependency_4}
    Case example of a BRAID output that failed QC although other dependencies passed.
    The preprocessed scan appears good quality, with reasonable values for estimated motion,
    number of detected outlier slices, and chi-squared analysis between the scan and signal reconstructed from the tensor fit.
    The FA map from the tensor fit also passed visual quality control.
    However, the output from BRAID shows misalignment of the tensor map with the MNI template.
    }
\end{figure}

A key finding from the aggregated QC outcomes is that Dependency Failed; Outcome Passed cases represent
a non-trivial proportion of results across all pipelines with upstream dependencies.
This pattern has important implications: the apparent quality of a derived output may not be sufficient to establish its trustworthiness.
A downstream map that passes visual QC may be compromised if an earlier stage produced erroneous outputs,
and this latent failure would go undetected without systematic inspection of the full pipeline.
To illustrate this concretely, we highlight four representative case examples
(Figs.~\ref{fig:dependency_1},~\ref{fig:dependency_2},~\ref{fig:dependency_3},~\ref{fig:dependency_4}).
This underscores the necessity of propagating QC assessments through the entire pipeline hierarchy rather than evaluating outputs in isolation.
Pipelines with more complex or numerous upstream dependencies tend to exhibit higher overall
failure rates (combining Dependency Passed; Outcome Failed, Dependency Failed; Outcome Passed, and Both Failed categories),
reflecting the cumulative probability of error accumulation across tightly coupled processing chains.

\subsection*{Algorithms with the same dependencies can exhibit different failure sensitivity}

Although two pipelines may share an identical set of upstream dependencies, the detectability of dependency failures from their own
outputs can differ substantially depending on how each algorithm utilizes its inputs.
Tractseg and BRAID provide an illustration of this: both algorithms depend on the outputs of PreQual, tensor fitting, and atlas registration,
yet their failure distributions differ markedly.
Tractseg relies heavily on the preprocessed DWI data from PreQual to compute fiber orientation distributions~(FODs),
which are subsequently used in conjunction with derived segmentation masks to guide tractography.
This multi-stage consumption of the preprocessed data means that failures introduced during preprocessing are more likely to be exposed in the tractography outputs,
making dependency failures relatively apparent during Tractseg QC.
BRAID, by contrast, uses the FA and MD map for registration purposes.
Because failures such as an improper synthetic image from Synb0-DisCo (Supplementary Figs.~\ref{fig:supp1},~\ref{fig:supp2})
are not always visually apparent in FA and MD maps, those failures affecting BRAID can only be reliably identified through the QC of the dependency stages directly.
This further underscores the necessity of propagating QC assessments through the entire pipeline hierarchy.

\subsection*{Spatial granularity of QC should reflect algorithm-specific output structure}

Unlike scalar summary statistics, many diffusion MRI derivatives are spatially structured,
and the distinction between local and global failure modes is algorithmically meaningful.
For Tractseg, which segments 72 anatomically distinct white matter bundles,
QC failures are often tract-specific: individual bundles may be wispy or absent while others appear intact.
Evaluating Tractseg at the level of individual tracts provides greater resolution than a single global pass/fail judgment,
and enables downstream analyses to selectively exclude affected bundles while retaining usable data.
In contrast, for algorithms such as atlas registration, a spatially localized failure may compromise the validity of the entire output and warrants global rejection.
These considerations suggest that QC granularity should be matched to the spatial structure of the algorithm's outputs
and to the intended downstream use of those outputs.

\section{Conclusion}

This work demonstrates the feasibility and analytical value of structured QC for diffusion MRI processing pipelines.
Across 18{,}328 scans and seven pipelines, we identified failure modes that aggregate metrics alone cannot surface.
A non-trivial proportion of visually acceptable outputs relied on failed upstream stages, establishing that QC must cover the full pipeline hierarchy.
The detectability of upstream failures also varies by algorithm.
Finally, QC granularity should reflect the spatial structure of each algorithm's outputs.
These results highlight the need for systematic, hierarchy-aware, and algorithm-specific QC in large-scale diffusion MRI research.


\subsection*{Disclosures}
Timothy J. Hohman is Deputy Editor for Alzheimer's \& Dementia: TRIC, Section Editor for Alzheimer's \& Dementia, and on the scientific advisory board for Circular Genomics.
All other authors declare that there are no financial interests, commercial affiliations, or other potential conflicts of interest that could have influenced the objectivity of this research or the writing of this paper.

\subsection*{Code, Data, and Materials Availability}
The source code for the QC app is available at \href{https://github.com/MASILab/ADSP_AutoQA}{\linkable{https://github.com/MASILab/ADSP\_AutoQA}}.
The datasets supporting the conclusions of this research are accessible under institutional data use agreements (DUAs) administered by their respective custodians.
ADNI data are accessible through the ADNI data sharing portal (adni.loni.usc.edu).
ROSMAPMARS data are accessible through the Rush Alzheimer's Disease Center Research Resource Sharing Hub (radc.rush.edu).
WASHU data are accessible through the Central Neuroimaging Data Archive (cnda.wustl.edu) under the Knight ADRC.
SCAN data are shared through the National Alzheimer's Coordinating Center (NACC).
NACC, FloridaADRC, and Indiana data are contributed by NIA-funded Alzheimer's Disease Research Centers and are governed by DUAs administered through those centers.
WRAP data are accessible through the Wisconsin Registry for Alzheimer's Prevention (\url{https://wrap.wisc.edu/data-requests-2/}).
HABS-HD data are accessible through the Institute for Translational Research at the University of North Texas Health Science Center (\url{https://apps.unthsc.edu/itr/reports}).

\subsection*{Acknowledgments}
This work was supported in part by the National Institutes of Health through awards K01-EB032898,
K01-AG073584, 1R01EB017230-01A1, U24~AG074855, U01~AG068057, and R01~AG059716,
and by the Alzheimer's Disease Sequencing Project Phenotype Harmonization Consortium~(ADSP-PHC).
Computing resources were provided by the Advanced Computing Center for Research and Education~(ACCRE) at Vanderbilt University.

Data collection and sharing for ADNI were supported by National Institutes of Health Grant U01-AG024904
and Department of Defense (award number W81XWH-12-2-0012).
ADNI is also funded by the National Institute on Aging, the National Institute of Biomedical
Imaging and Bioengineering, and through generous contributions from the following:
AbbVie, Alzheimer’s Association; Alzheimer’s Drug Discovery Foundation; Araclon Biotech; BioClinica, Inc.;
Biogen; Bristol-Myers Squibb Company; CereSpir, Inc.; Cogstate; Eisai Inc.; Elan Pharmaceuticals, Inc.;
Eli Lilly and Company; EuroImmun; F. Hoffmann-La Roche Ltd and its affiliated company Genentech, Inc.;
Fujirebio; GE Healthcare; IXICO Ltd.; Janssen Alzheimer Immunotherapy Research \& Development, LLC.;
Johnson \& Johnson Pharmaceutical Research \& Development LLC.; Lumosity; Lundbeck; Merck \& Co., Inc.;
Meso Scale Diagnostics, LLC.; NeuroRx Research; Neurotrack Technologies; Novartis Pharmaceuticals Corporation;
Pfizer Inc.; Piramal Imaging; Servier; Takeda Pharmaceutical Company; and Transition Therapeutics.
The Canadian Institutes of Health Research is providing funds to support ADNI clinical sites in Canada.
Private sector contributions are facilitated by the Foundation for the National Institutes of Health (www.fnih.org).
The grantee organization is the Northern California Institute for Research and Education,
and the study is coordinated by the Alzheimer’s Therapeutic Research Institute at the University of Southern California.
ADNI data are disseminated by the Laboratory for Neuro Imaging at the University of Southern California.
Data used in the preparation of this article were obtained from the Alzheimer’s Disease Neuroimaging
Initiative (ADNI) database (adni.loni.usc.edu). The ADNI was launched in 2003 as a public private partnership,
led by Principal Investigator Michael W. Weiner, MD.
The original goal of ADNI was to test whether serial magnetic resonance imaging (MRI),
positron emission tomography (PET), other biological markers,
and clinical and neuropsychological assessment can be combined to measure the progression
of mild cognitive impairment (MCI) and early Alzheimer’s disease (AD).
The current goals include validating biomarkers for clinical trials,
improving the generalizability of ADNI data by increasing diversity in the participant cohort,
and to provide data concerning the diagnosis and progression of Alzheimer’s disease to the scientific community.
For up-to-date information, see \url{https://adni.loni.usc.edu}.

Research reported in this publication as a part of the Health and Aging Brain Study:
Health Disparities (HABS-HD) was supported by the National Institute on Aging of the National Institutes
of Health under Award Numbers R01AG054073 and R01AG058533, R01AG070862, P41EB015922 and U19AG078109.
The content is solely the responsibility of the authors and does not necessarily represent the official
views of the National Institutes of Health.

The NACC database is funded by NIA/NIH Grant U24 AG072122.
NACC data are contributed by the NIA-funded ADRCs: P30 AG062429 (PI James Brewer, MD, PhD),
P30 AG066468 (PI Oscar Lopez, MD), P30 AG062421 (PI Bradley Hyman, MD, PhD),
P30 AG066509 (PI Thomas Grabowski, MD), P30 AG066514 (PI Mary Sano, PhD),
P30 AG066530 (PI Helena Chui, MD), P30 AG066507 (PI Marilyn Albert, PhD),
P30 AG066444 (PI John Morris, MD), P30 AG066518 (PI Jeffrey Kaye, MD), P30 AG066512 (PI Thomas Wisniewski, MD),
P30 AG066462 (PI Scott Small, MD), P30 AG072979 (PI David Wolk, MD), P30 AG072972 (PI Charles DeCarli, MD),
P30 AG072976 (PI Andrew Saykin, PsyD), P30 AG072975 (PI David Bennett, MD), P30 AG072978 (PI Ann McKee, MD),
P30 AG072977 (PI Robert Vassar, PhD), P30 AG066519 (PI Frank LaFerla, PhD),
P30 AG062677 (PI Ronald Petersen, MD, PhD), P30 AG079280 (PI Eric Reiman, MD),
P30 AG062422 (PI Gil Rabinovici, MD), P30 AG066511 (PI Allan Levey, MD, PhD),
P30 AG072946 (PI Linda Van Eldik, PhD), P30 AG062715 (PI Sanjay Asthana, MD, FRCP),
P30 AG072973 (PI Russell Swerdlow, MD), P30 AG066506 (PI Todd Golde, MD, PhD),
P30 AG066508 (PI Stephen Strittmatter, MD, PhD), P30 AG066515 (PI Victor Henderson, MD, MS),
P30 AG072947 (PI Suzanne Craft, PhD), P30 AG072931 (PI Henry Paulson, MD, PhD),
P30 AG066546 (PI Sudha Seshadri, MD), P20 AG068024 (PI Erik Roberson, MD, PhD),
P20 AG068053 (PI Justin Miller, PhD), P20 AG068077 (PI Gary Rosenberg, MD),
P20 AG068082 (PI Angela Jefferson, PhD), P30 AG072958 (PI Heather Whitson, MD),
P30 AG072959 (PI James Leverenz, MD).

Data contributed from MAP/ROS/MARS was supported by NIA R01AG017917, P30AG10161, P30AG072975,
R01AG022018, R01AG056405, UH2NS100599, UH3NS100599, R01AG064233, R01AG15819 and R01AG067482,
and the Illinois Department of Public Health (Alzheimer’s Disease Research Fund).
Data can be accessed at \url{https://www.radc.rush.edu}.
More information about participant demographics and study information can be found here:
\url{https://www.rushu.rush.edu/research-rush-university/departmental-research/rush-alzheimers-disease-center/rush-alzheimers-disease-center-research/epidemiologic-research}.

The data contributed from the Wisconsin Registry for Alzheimer’s Prevention was supported by NIA AG021155,
AG0271761, AG037639, and AG054047.

We thank Knight ADRC for providing neuroimaging data to us (WASHU dataset).
The data contributed through WASHU (Knight ADRC) was supported by grant numbers P30 AG066444,
P01 AG03991, and P01 AG026276. For WASHU, Clinical Dementia Ratings (CDRs) are obtained
from assessments by experienced clinicians trained in the use of the CDR.

The NACC database is funded by NIA/NIH Grant U24 AG072122.
SCAN is a multi-institutional project that was funded as a U24 grant (AG067418)
by the National Institute on Aging in May 2020. Data collected by SCAN and shared by NACC are contributed
by the NIA-funded ADRCs as follows:
Arizona Alzheimer’s Center - P30 AG072980 (PI: Eric Reiman, MD);
R01 AG069453 (PI: Eric Reiman (contact), MD); P30 AG019610 (PI: Eric Reiman, MD);
and the State of Arizona which provided additional funding supporting our center;
Boston University - P30 AG013846 (PI Neil Kowall MD); Cleveland ADRC - P30 AG062428 (James Leverenz, MD);
Cleveland Clinic, Las Vegas – P20AG068053; Columbia - P50 AG008702 (PI Scott Small MD);
Duke/UNC ADRC – P30 AG072958; Emory University - P30AG066511 (PI Levey Allan, MD, PhD);
Indiana University - R01 AG19771 (PI Andrew Saykin, PsyD); P30 AG10133 (PI Andrew Saykin, PsyD);
P30 AG072976 (PI Andrew Saykin, PsyD); R01 AG061788 (PI Shannon Risacher, PhD);
R01 AG053993 (PI Yu-Chien Wu, MD, PhD); U01 AG057195 (PI Liana Apostolova, MD);
U19 AG063911 (PI Bradley Boeve, MD); and the Indiana University Department of Radiology and Imaging Sciences;
Johns Hopkins - P30 AG066507 (PI Marilyn Albert, Phd.); Mayo Clinic - P50 AG016574 (PI Ronald Petersen MD PhD);
Mount Sinai - P30 AG066514 (PI Mary Sano, PhD); R01 AG054110 (PI Trey Hedden, PhD);
R01 AG053509 (PI Trey Hedden, PhD); New York University - P30AG066512-01S2 (PI Thomas Wisniewski, MD);
R01AG056031 (PI Ricardo Osorio, MD); R01AG056531 (PIs Ricardo Osorio, MD; Girardin Jean-Louis, PhD);
Northwestern University - P30 AG013854 (PI Robert Vassar PhD); R01 AG045571 (PI Emily Rogalski, PhD);
R56 AG045571, (PI Emily Rogalski, PhD); R01 AG067781, (PI Emily Rogalski, PhD);
U19 AG073153, (PI Emily Rogalski, PhD); R01 DC008552, (M.-Marsel Mesulam, MD);
R01 AG077444, (PIs M.-Marsel Mesulam, MD, Emily Rogalski, PhD);
R01 NS075075 (PI Emily Rogalski, PhD); R01 AG056258 (PI Emily Rogalski, PhD);
Oregon Health and Science University - P30 AG008017 (PI Jeffrey Kaye MD);
R56 AG074321 (PI Jeffrey Kaye, MD); Rush University - P30 AG010161 (PI David Bennett MD);
Stanford – P30AG066515; P50 AG047366 (PI Victor Henderson MD MS);
University of Alabama, Birmingham – P20; University of California, Davis - P30 AG10129 (PI Charles DeCarli, MD);
P30 AG072972 (PI Charles DeCarli, MD); University of California, Irvine - P50 AG016573 (PI Frank LaFerla PhD);
University of California, San Diego - P30AG062429 (PI James Brewer, MD, PhD);
University of California, San Francisco - P30 AG062422 (Rabinovici, Gil D., MD);
University of Kansas - P30 AG035982 (Russell Swerdlow, MD);
University of Kentucky - P30 AG028283-15S1 (PIs Linda Van Eldik, PhD and Brian Gold, PhD);
University of Michigan ADRC - P30AG053760 (PI Henry Paulson, MD, PhD) P30AG072931 (PI Henry Paulson, MD, PhD)
Cure Alzheimer’s Fund 200775 - (PI Henry Paulson, MD, PhD) U19 NS120384
(PI Charles DeCarli, MD, University of Michigan Site PI Henry Paulson, MD, PhD) R01 AG068338
(MPI Bruno Giordani, PhD, Carol Persad, PhD, Yi Murphey, PhD) S10OD026738-01 (PI Douglas Noll, PhD)
R01 AG058724 (PI Benjamin Hampstead, PhD) R35 AG072262 (PI Benjamin Hampstead, PhD) W81XWH2110743 (PI Benjamin Hampstead, PhD)
R01 AG073235 (PI Nancy Chiaravalloti, University of Michigan Site PI Benjamin Hampstead, PhD)
1I01RX001534 (PI Benjamin Hampstead, PhD) IRX001381 (PI Benjamin Hampstead, PhD);
University of New Mexico - P20 AG068077 (Gary Rosenberg, MD);
University of Pennsylvania - State of PA project 2019NF4100087335 (PI David Wolk, MD);
Rooney Family Research Fund (PI David Wolk, MD); R01 AG055005 (PI David Wolk, MD);
University of Pittsburgh - P50 AG005133 (PI Oscar Lopez MD); University of Southern California - P50 AG005142 (PI Helena Chui MD);
University of Washington - P50 AG005136 (PI Thomas Grabowski MD);
University of Wisconsin - P50 AG033514 (PI Sanjay Asthana MD FRCP);
Vanderbilt University – P20 AG068082; Wake Forest - P30AG072947 (PI Suzanne Craft, PhD);
Washington University, St. Louis - P01 AG03991 (PI John Morris MD); P01 AG026276 (PI John Morris MD);
P20 MH071616 (PI Dan Marcus); P30 AG066444 (PI John Morris MD); P30 NS098577 (PI Dan Marcus);
R01 AG021910 (PI Randy Buckner); R01 AG043434 (PI Catherine Roe); R01 EB009352 (PI Dan Marcus);
UL1 TR000448 (PI Brad Evanoff); U24 RR021382 (PI Bruce Rosen); Avid Radiopharmaceuticals / Eli Lilly;
Yale - P50 AG047270 (PI Stephen Strittmatter MD PhD); R01AG052560 (MPI: Christopher van Dyck, MD;
Richard Carson, PhD); R01AG062276 (PI: Christopher van Dyck, MD); 1Florida - P30AG066506-03 (PI Glenn Smith, PhD);
P50 AG047266 (PI Todd Golde MD PhD)

Source (DICOM) and derived data is available for access via web download from
the Central Neuroimaging Data Archive (CNDA) \url{cnda.wustl.edu} upon request from Knight-ADRC:
\url{https://knightadrc.wustl.edu/professionals-clinicians/request-center-resources/submit-a-request/}.

We have used AI as a tool in the creation of this content,
however, the foundational ideas, underlying concepts, and original gist stem directly from the
personal insights, creativity, and intellectual effort of the author(s).
The use of generative AI serves to enhance and support the author's original contributions
by assisting in the ideation, drafting, and refinement processes.
All AI-assisted content has been carefully reviewed, edited, and approved by the author(s)
to ensure it aligns with the intended message, values, and creativity of the work.


\bibliography{report}
\bibliographystyle{spiejour}

\clearpage

\begin{center}
{\LARGE\textbf{Supplemental Material}}
\end{center}
\vspace{1em}

\setcounter{figure}{0}
\renewcommand{\thefigure}{S\arabic{figure}}
\renewcommand{\theHfigure}{S\arabic{figure}}

\setcounter{table}{0}
\renewcommand{\thetable}{S\arabic{table}}
\renewcommand{\theHtable}{S\arabic{table}}

\setcounter{equation}{0}
\renewcommand{\theequation}{S\arabic{equation}}
\renewcommand{\theHequation}{S\arabic{equation}}

\section{Additional Pipeline QC Procedures}

\subsection{Brain Shape Computing Toolbox}

\begin{figure}[ht]
\begin{center}
\begin{tabular}{c}
\includegraphics[width=\linewidth]{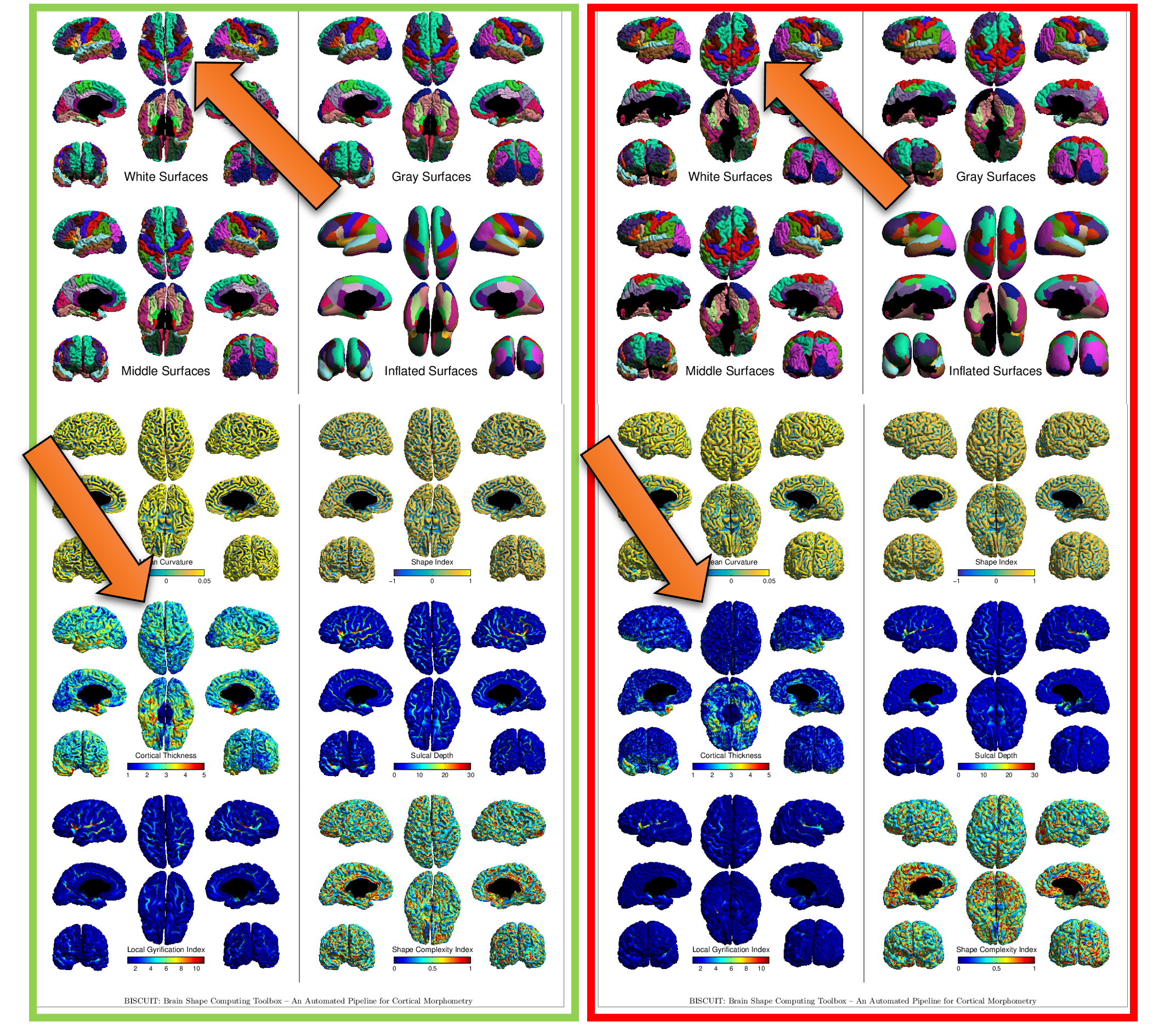}
\end{tabular}
\end{center}
\caption{\label{fig:supp3}
(Left) The BISCUIT Toolbox outputs are deemed successful if the cortical surface parcellation appears symmetric across hemispheres and the cortical thickness measurements hover around the 2 to 4~mm range. (Right) Most common failures are when the cortical parcellation is not symmetric or the cortical thickness measurements are suspiciously low.}
\end{figure}

Brain Shape Computing Toolbox (BISCUIT) QC consists of visualizing the cortical surface reconstruction parcellation
as well as various cortical surface scalars mapped onto the brain surface.
We expect the parcellation of the cortex to be mostly symmetric when comparing hemispheres,
with segmented regions in the proper positions (Supplementary Fig.~\ref{fig:supp3}).
The most indicative quantitative scalar map for QC is cortical thickness, which should be in a range of around 2--4 millimeters on average.
For failed instances, the parcellation is either visibly asymmetric, the regions are not in their proper locations,
or some regions are missing.
Most failed outputs have cortical thickness maps that are much lower than the expected range.

\subsection{Freesurfer}

\begin{figure}[ht]
\begin{center}
\begin{tabular}{c}
\includegraphics[width=\linewidth]{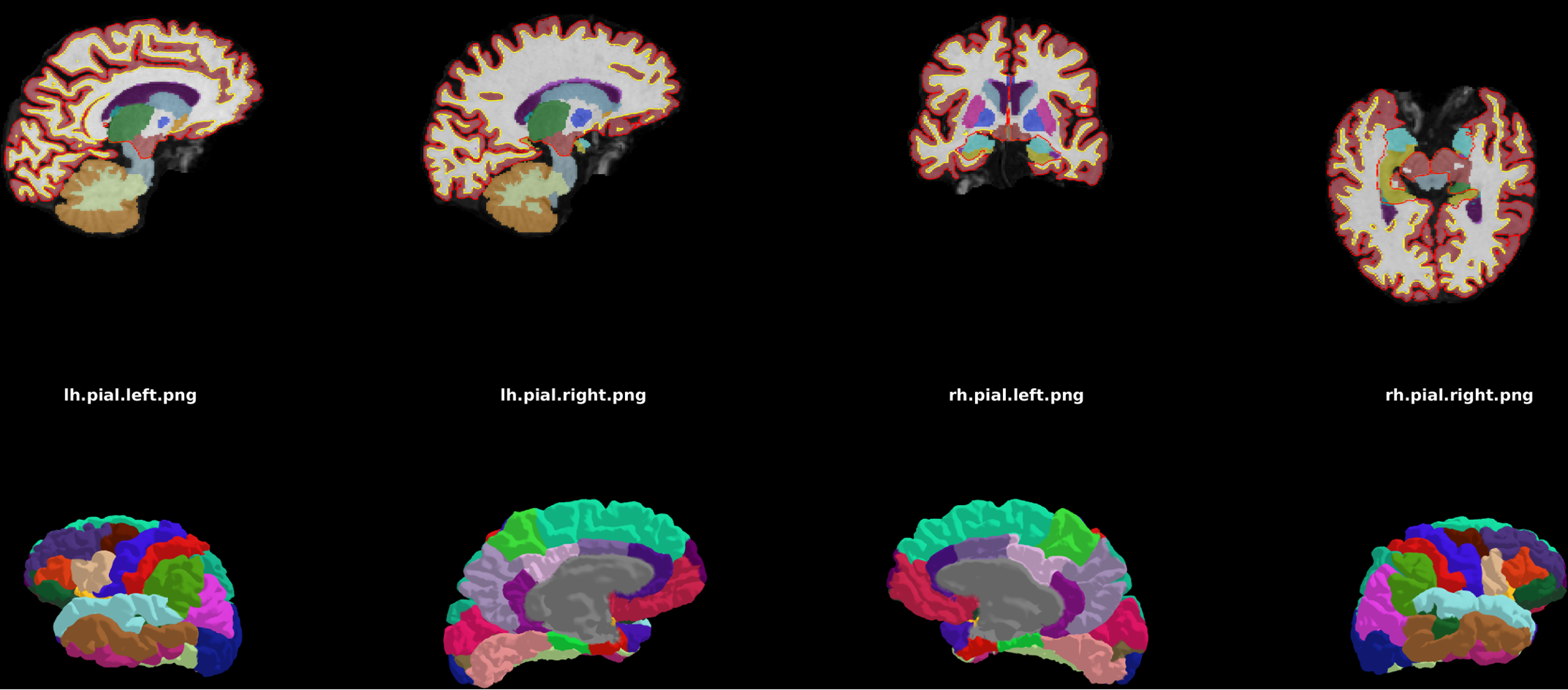}
\end{tabular}
\end{center}
\caption{\label{fig:supp4}
We expect the brain parcellation of Freesurfer to properly segment ROIs and define accurate CSF/WM/GM boundaries. For most pipeline instances, failure to generate a proper Freesurfer QC PNG is the main indicator that the algorithm output is a failure.}
\end{figure}

Freesurfer QC visualization is an overlay of the segmentation maps and gray matter surface contour lines on the T1w image that was used as input.
We expect the segmentations to align with their corresponding structures, for instance,
the ventricles, the white matter map, the cerebellum, etc.\ (Supplementary Fig.~\ref{fig:supp4}).
The boundary between the gray and white matter segmentations should also be near the cortex.
In most instances when the Freesurfer algorithm is unsuccessful, it fails to generate all the outputs, so the QC PNG cannot be made.

\subsection{NODDI}

For NODDI, we visualize axial slices for the estimated NODDI scalar measures of orientation dispersion index (ODI),
isotropic volume fraction~(ISOVF), intracellular volume fraction~(ICVF), as well as an FA map for reference of white matter tracts.
A high ODI value means that the orientations of axons are more disperse.
Thus, very directed white matter tracts such as the corpus callosum should be dark.
As ISOVF represents the fraction of cerebrospinal fluid for a voxel, we expect the CSF to be very bright.
Finally, the ICVF is the fraction of voxels occupied by intracellular space, or axons and dendrites (collectively, neurites),
as opposed to glial cells and cell bodies, so we expect the values in white matter to be larger than in the gray matter.
Most failures are obvious, with no anatomical structures visible on the NODDI maps.

\subsection{Nextflow Bundle Analysis}

\begin{figure}[ht]
\begin{center}
\begin{tabular}{c}
\includegraphics[width=\linewidth]{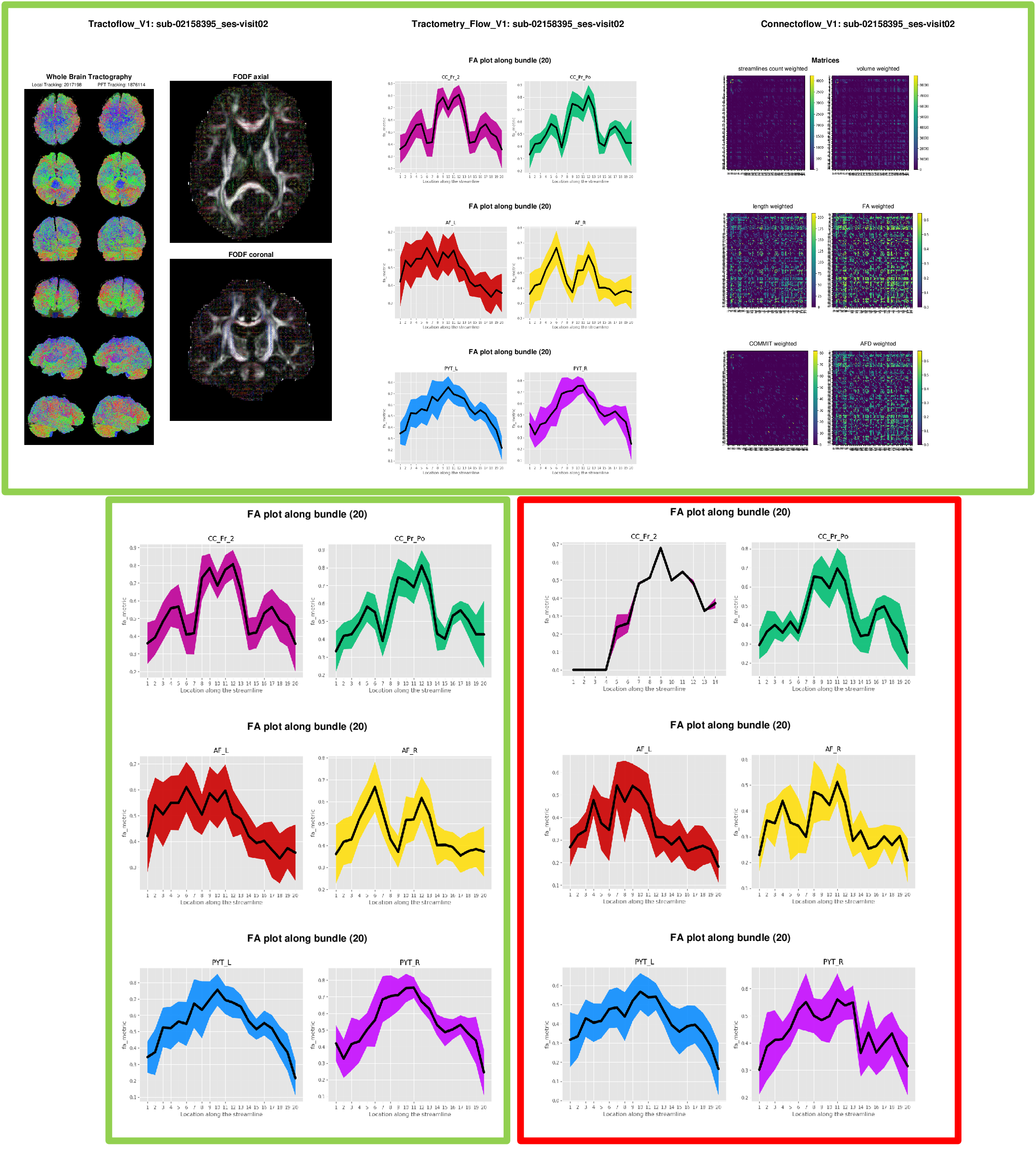}
\end{tabular}
\end{center}
\caption{\label{fig:supp6}
(top) We expect the whole brain tractogram to have streamlines extending to all parts of the brain, connectomes to have a reasonable number of streamlines, and (bottom, green) the mean FA along tracts to be relatively higher values. For most failure instances, the pipeline does not output a QC PDF file. (bottom, red) There are a few instances when the mean FA along tracts is not consistent with expectations, but we have found these occurrences to be rare}
\end{figure}

For Nextflow Bundle Analysis,
we visualize the whole brain tractography, the mean FA along the corpus callosum, arcuate fasciculus,
and the corticospinal tract (also known as the pyramidal tract) with standard deviation value,
and connectomics matrices weighted by various quantitative metrics.
The whole-brain tractogram is expected to have streamlines reaching every part of the cortex in the brain,
with a large crossing fiber that connects the two brain hemispheres. We expect mean FA values along segmented bundles to range from 0.4 to 0.9,
with values centered on average around 0.5 (and slightly higher for the corpus callosum) (Supplementary Fig.~\ref{fig:supp6}).
The connectomics matrices, like with the Connectome Special, should have a reasonable number of connections.
Like Freesurfer, in most instances when the algorithm is unsuccessful, it fails to generate a QC PDF.

\clearpage
\begin{figure}[ht]
\begin{center}
\begin{tabular}{c}
\includegraphics[width=\linewidth]{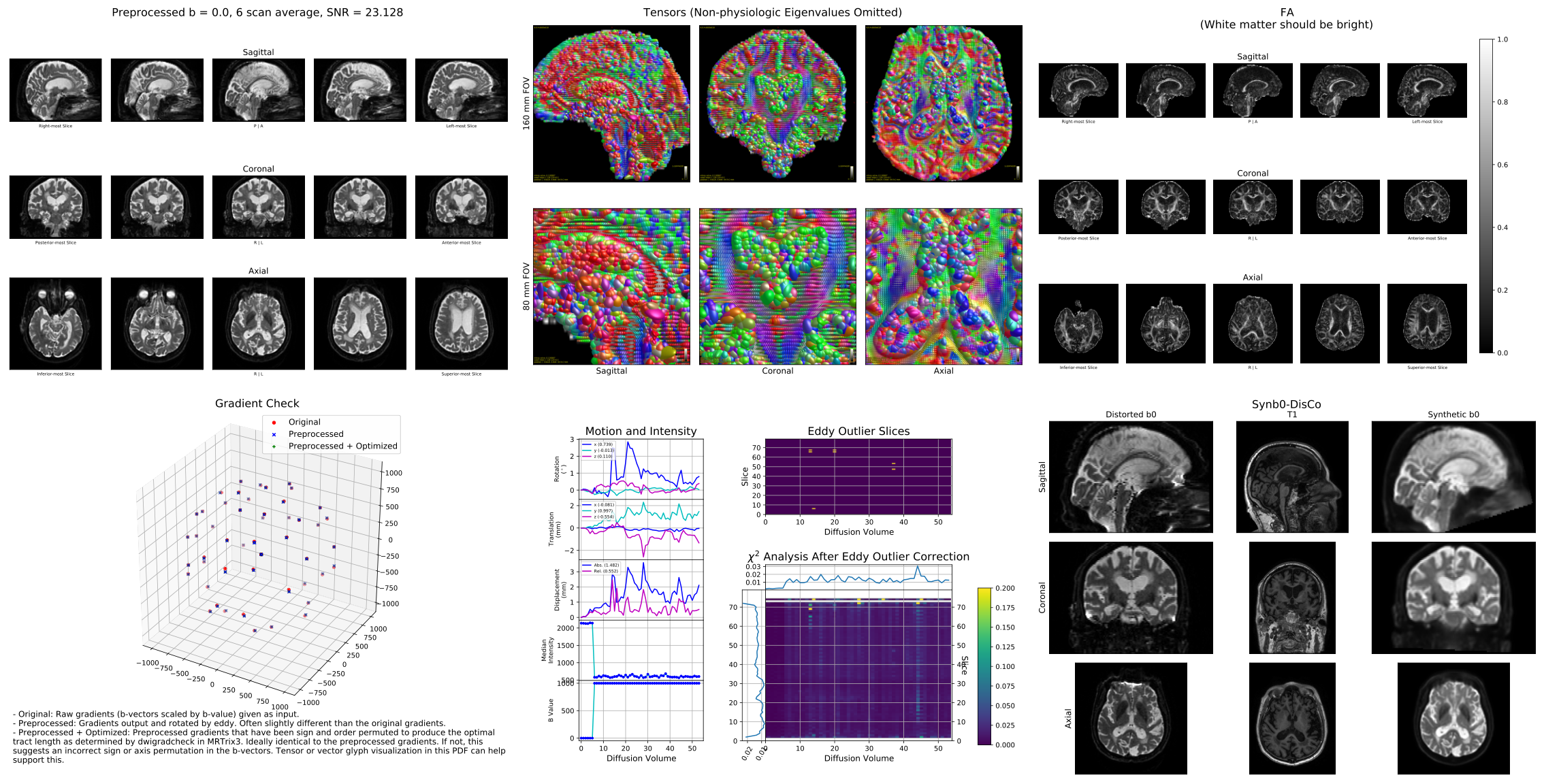}
\end{tabular}
\end{center}
\caption{\label{fig:prequal_pdf}
Example of a single image PNG output converted from the multi-page PreQual PDF.
On the top row are visualizations of the preprocessed data.
From top left to top right: slices of the preprocessed data (specifically all b0 scans averaged),
tensor maps of the brain, FA map.
On the bottom row are additional visualizations to ensure data quality.
From left to right, re-orientations of the supplied b-vectors,
motion/intensity/outlier charts along with a chi-squared map of the difference between the original image
intensity and the intensity from tensor reconstruction,
and the synthetic b0 for distortion correction generated from a supplied T1w image
(and compared to an original, unprocessed b0 for the DWI scan).
}
\end{figure}

\clearpage
\begin{figure}[ht]
\begin{center}
\begin{tabular}{c}
\includegraphics[width=\linewidth]{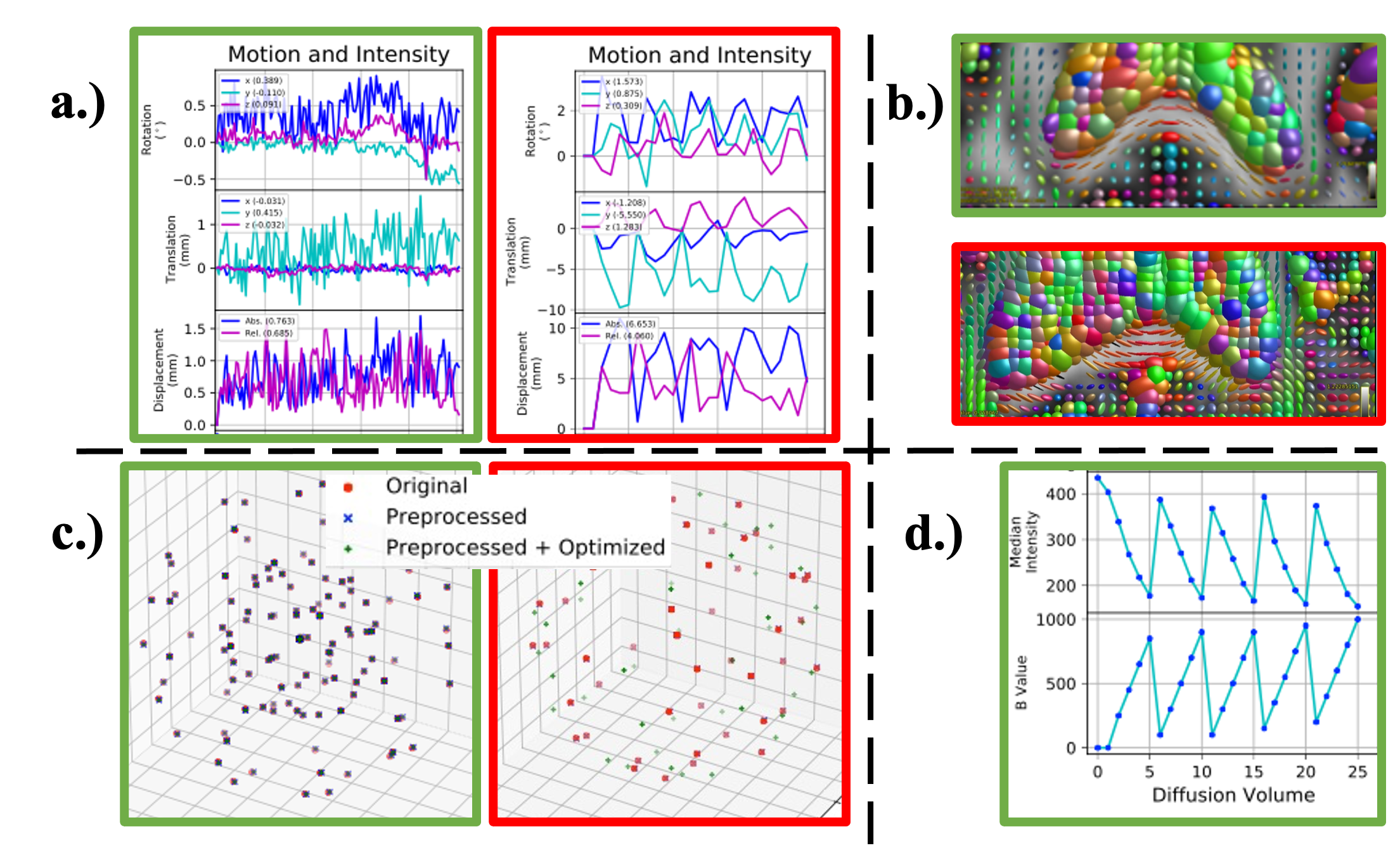}
\end{tabular}
\end{center}
\caption{\label{fig:prequal}
Representative PreQual QC example.
(a., green) While motion in DWI scans is not uncommon, we expect fluctuations between adjacent volumes to be
relatively small. (a., red) However, an indication that the EDDY algorithm has failed is when intervolume motion is
consistently on the magnitude of several millimeters. Properly oriented b-vector files will yield tensors whose
directions point along white matter tracts, such as (b., green) the corpus callosum, and (c., green) will already have the
optimal orientation. Any orientation issues will result in (b., red) the tensors pointing in the wrong direction or (c., red)
will be determined as suboptimal when permuting the axes. (d.) The expectation for median volume intensity is roughly
exponentially decreasing with increasing b-values.
}
\end{figure}

\clearpage
\begin{figure}[ht]
\begin{center}
\begin{tabular}{c}
\includegraphics[width=\linewidth]{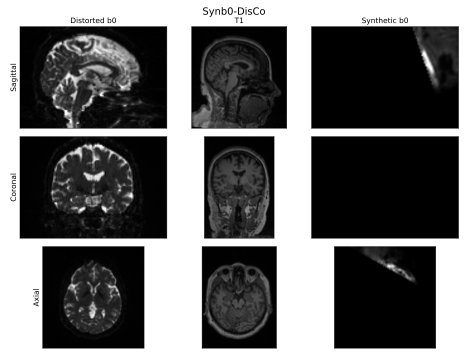}
\end{tabular}
\end{center}
\caption{\label{fig:supp1}
A common example of a Synb0-DisCo failure case, where the synthesized image is not at all representative of the original anatomy used as input to the algorithm.
}
\end{figure}

\clearpage
\begin{figure}[ht]
\begin{center}
\begin{tabular}{c}
\includegraphics[width=\linewidth]{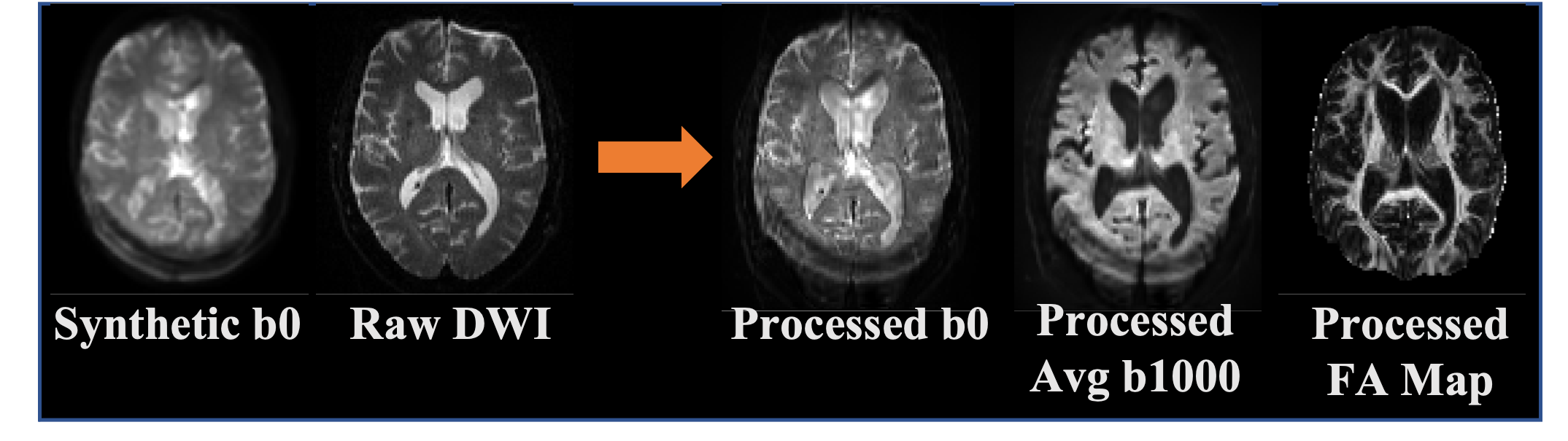}
\end{tabular}
\end{center}
\caption{\label{fig:supp2}
Not all algorithmic failures are large failures: some can be subtle to the point where it appears downstream analyses are not affected, highlighting the need to QC all parts of a given pipeline or series of pipelines. The given raw DWI scan and a synthetic b0 image, for which the algorithm has hallucinated an intense signal from the back of a skull, are fed as input to the TOPUP algorithm. While the resulting preprocessed DWI scan presents the same hallucinated effects, the derived FA map does not indicate any abnormalities with the scan.
}
\end{figure}

\clearpage
\begin{figure}[ht]
\begin{center}
\begin{tabular}{c}
\includegraphics[width=\linewidth]{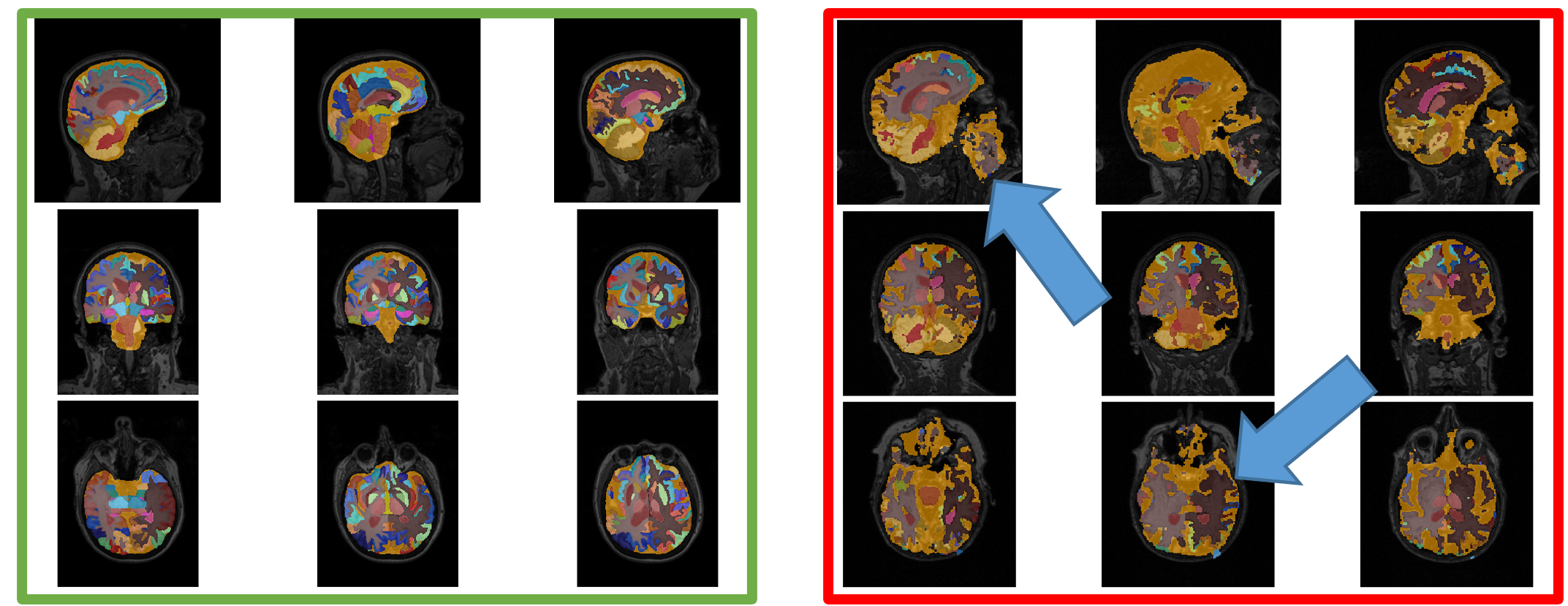}
\end{tabular}
\end{center}
\caption{\label{fig:segmentation}
Examples of SLANT-TICV QC output.
A good SLANT-TICV segmentation result has the TICV label (orange in figure) encapsulating the whole
intracranial vault, with the cortical substructures appropriately labeled as well. (Right) Most common improper
segmentations have labels outside of the brain or incorrectly segmented regions. MaCRUISE and UNesT QC documents
and processes are similar to those of SLANT-TICV.}
\end{figure}

\clearpage
\begin{figure}[ht]
\begin{center}
\begin{tabular}{c}
\includegraphics[width=\linewidth]{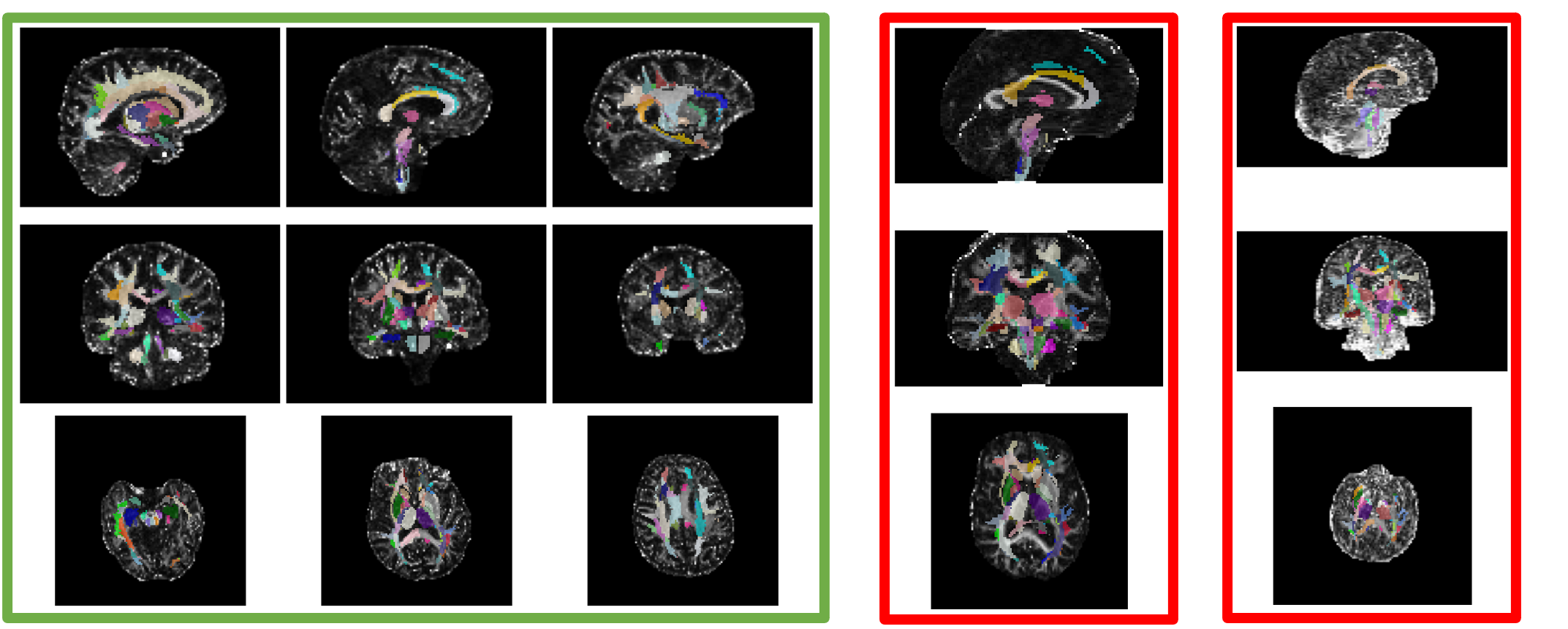}
\end{tabular}
\end{center}
\caption{\label{fig:eve3}
Examples of QC outputs for registration using the JHU template.
(Left) Proper registration of the EVE template into the DWI data space results in an alignment of segmented
white matter pathways from the annotated EVE3 label map. Additionally, a tensor fit is considered to pass QC if the
white matter tracts show as bright compared to the rest of the brain. (Middle) A poor registration can be seen when the
labels do not align with the white matter tracts, (Right) and a poor tensor fit can result in a very noisy FA map. All
images are viewed in the native participant DWI space.
}
\end{figure}

\clearpage
\begin{figure}[ht]
\begin{center}
\begin{tabular}{c}
\includegraphics[width=\linewidth]{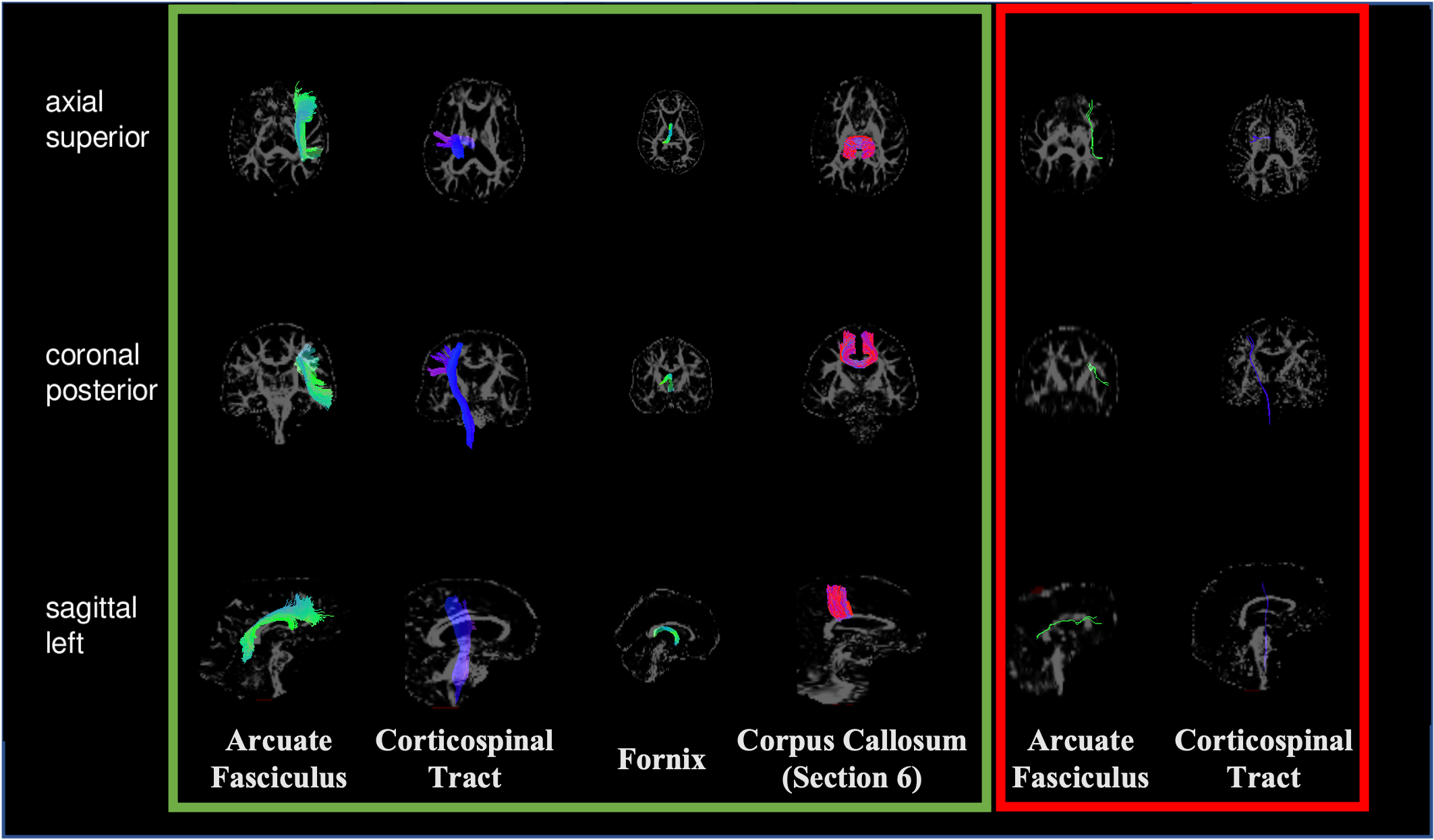}
\end{tabular}
\end{center}
\caption{\label{fig:tractseg}
Examples of Tractseg QC output.
When Tractseg bundles are properly generated, there are a sufficient number of streamlines to give the bundle a
full appearance (examples of major bundles in green). The most common failed outputs for Tractseg occur when there
are only a handful of streamlines, giving the bundle a wispy appearance upon visualization (examples in red).
}
\end{figure}

\clearpage

\clearpage

\clearpage
\begin{figure}[ht]
\begin{center}
\begin{tabular}{c}
\includegraphics[width=\linewidth]{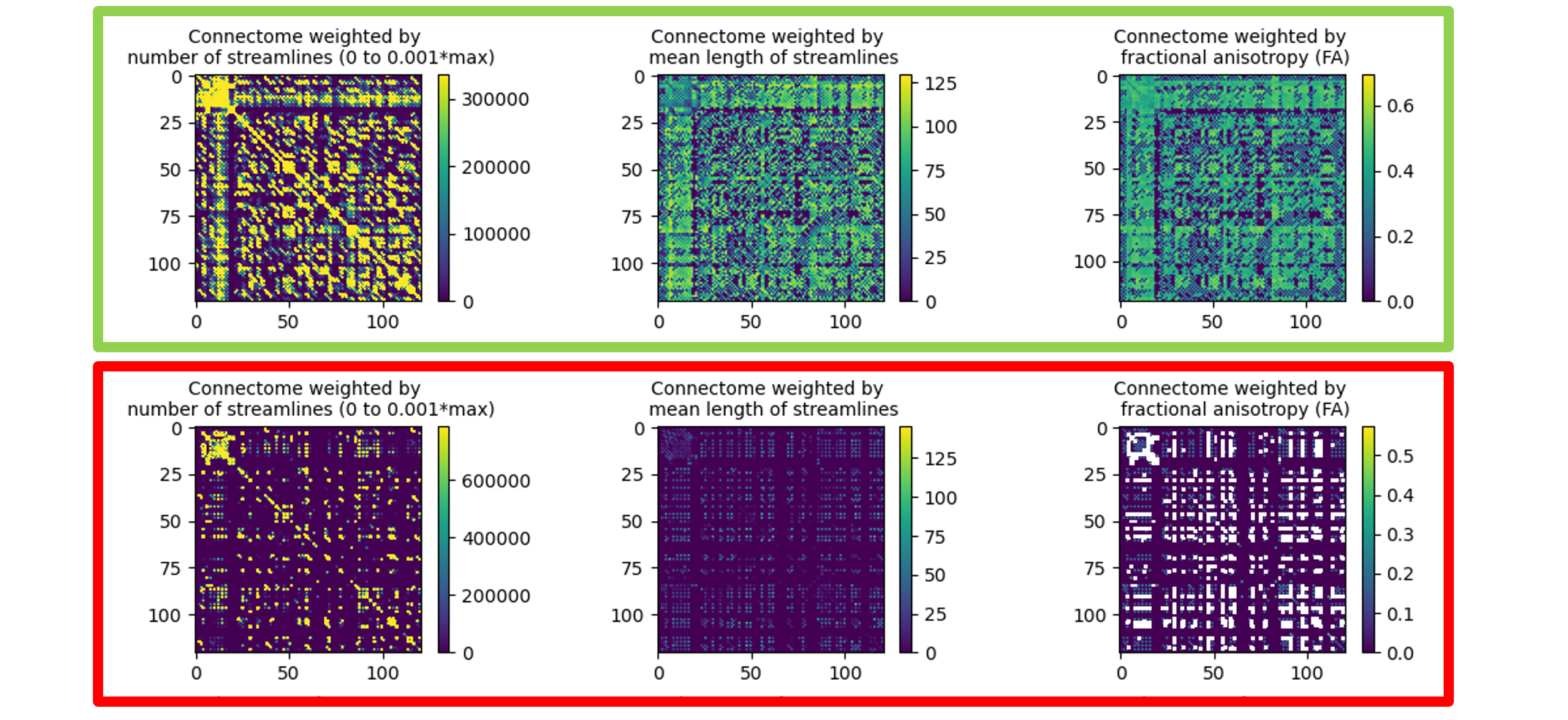}
\end{tabular}
\end{center}
\caption{\label{fig:connectome}
Examples of connectome special QC outputs.
(Top) For properly generated structural connectome matrices, we expect there to be a strong main diagonal
when weighted by the number of streamlines and a relatively homogeneous intensity for connectomes weighted by mean
FA. (Bottom) For failed outputs, the connectomes show sparse connectivity between brain regions, with many invalid
connections for mean FA weighting and a weak main diagonal when weighting by number of streamlines.
}
\end{figure}

\clearpage
\begin{figure}[ht]
\begin{center}
\begin{tabular}{c}
\includegraphics[width=\linewidth]{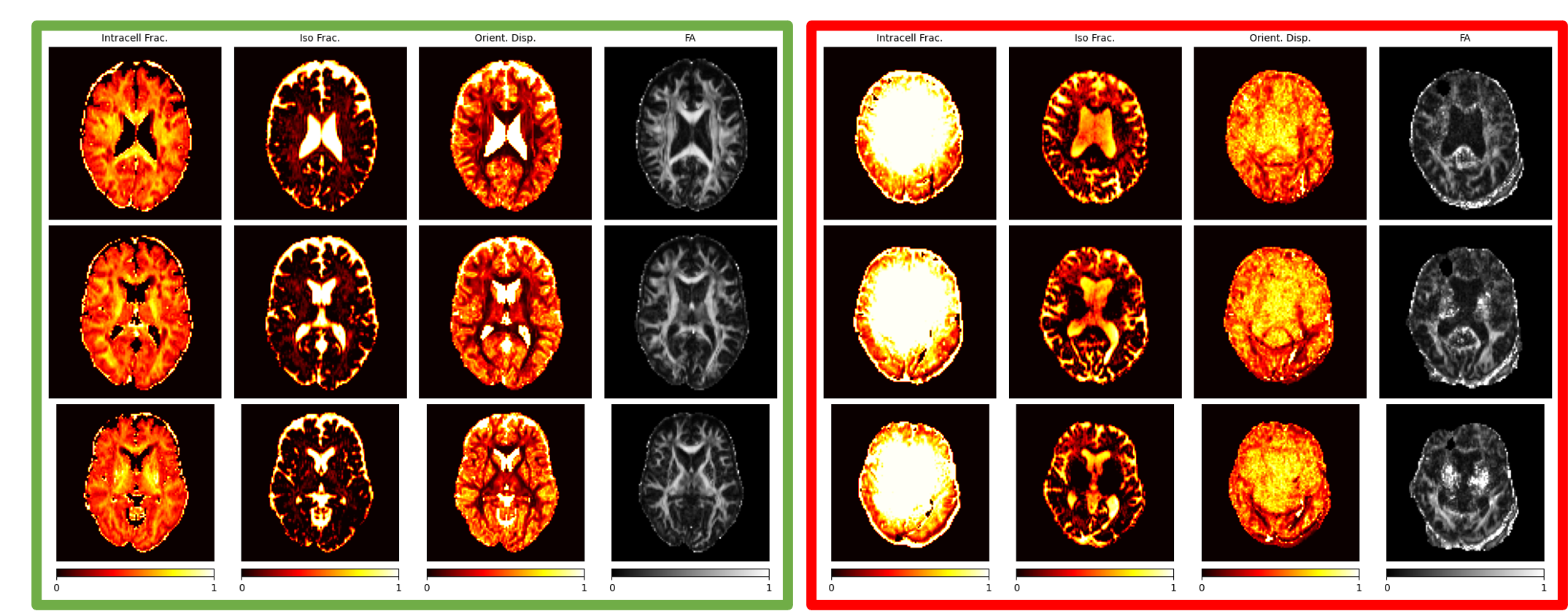}
\end{tabular}
\end{center}
\caption{\label{fig:connectome_fa}
(Left) For NODDI outputs, we expect large white matter pathways (visible as high intensity for the FA map) to
have a higher intracellular volume fraction, a low isotropic volume fraction, and a low orientation dispersion. In
contrast, the CSF should have an intracellular volume fraction of zero and both isotropic volume fraction and orientation
dispersion close to one. (Right) We consider NODDI outputs as failures if the estimated parameters do not adhere to
these expectations.
}
\end{figure}

\clearpage
\begin{figure}[ht]
\begin{center}
\begin{tabular}{c}
\includegraphics[width=\linewidth]{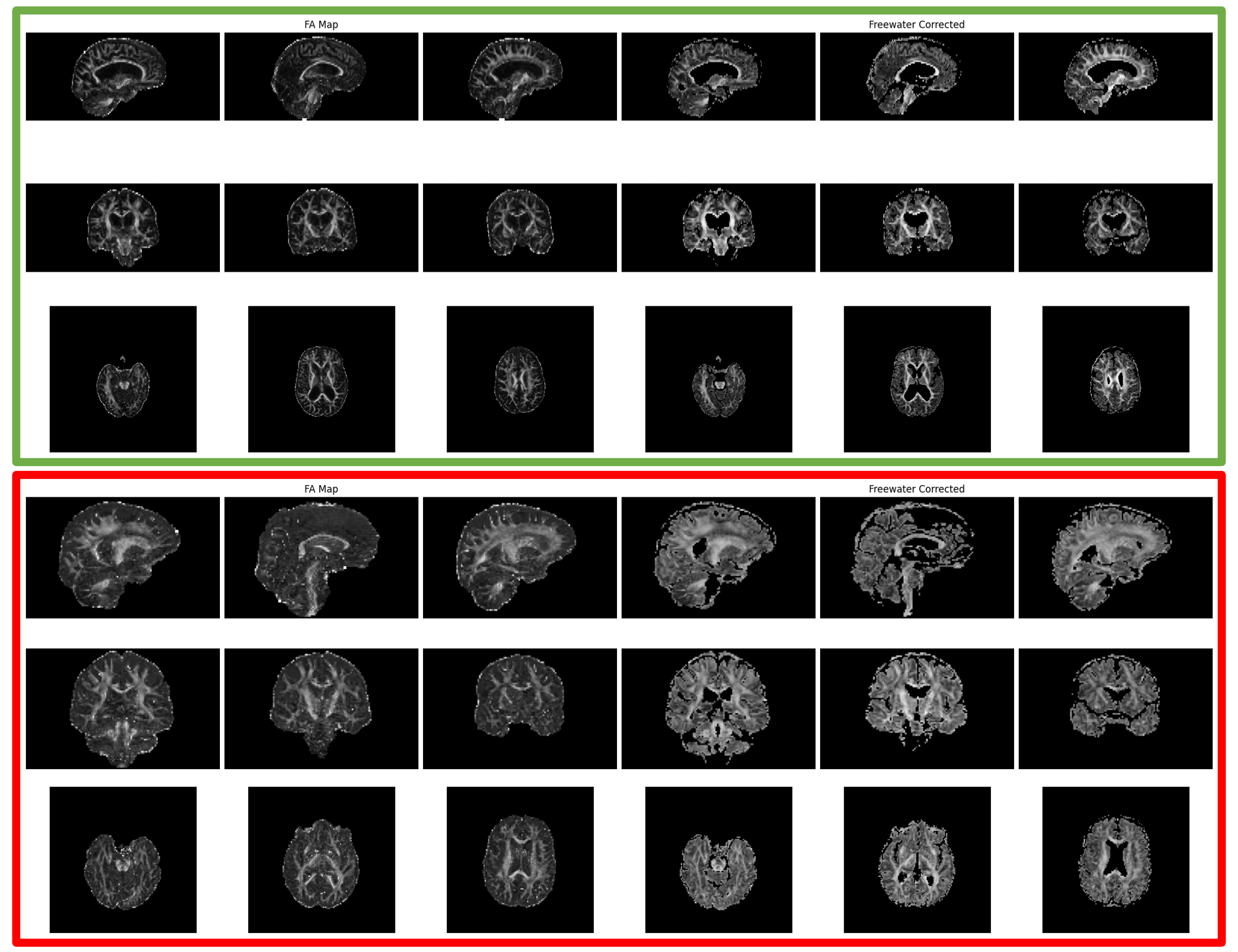}
\end{tabular}
\end{center}
\caption{\label{fig:supp5}
(top) For the Pasternak dual-tensor free water model fitting, we expect the WM tracts in the free water-corrected FA map to become enhanced when compared to the original image. (bottom) Most common failures occur when the dual-tensor model fit results in an overestimation of the FA in non-white matter brain regions.}
\end{figure}

\clearpage

\clearpage
\begin{figure}[ht]
\begin{center}
\begin{tabular}{c}
\includegraphics[width=\linewidth]{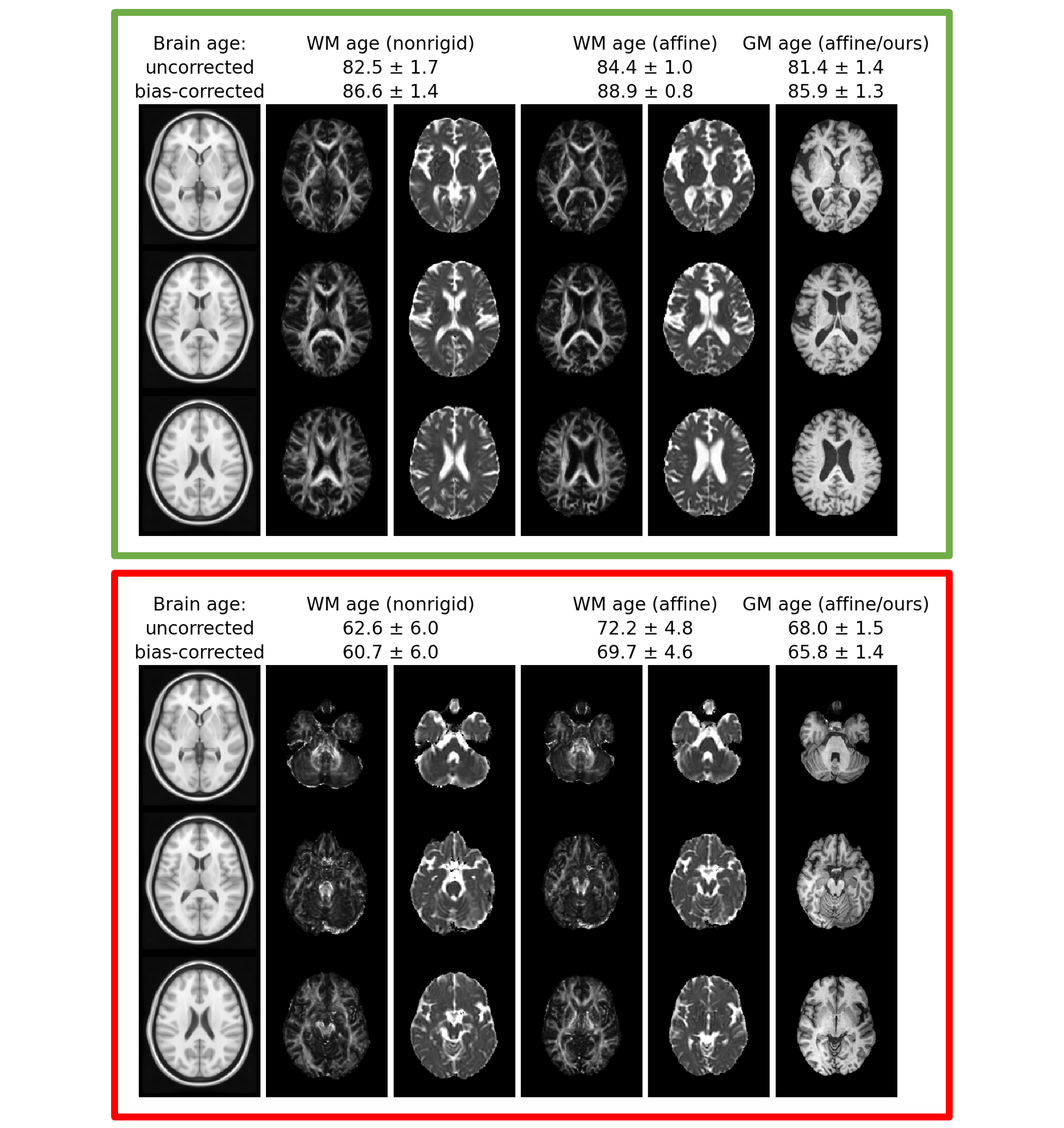}
\end{tabular}
\end{center}
\caption{\label{fig:supp7}
In the BRAID pipeline, we examine the FA/MD maps and the T1w scan that are registered to the MNI template (leftmost column), providing the estimated brain ages from the model outputs as well for quantitative quality assurance. (top) We expect the non-linear deformation (2nd and 3rd leftmost columns) to warp the brain so that macrostructural information is minimized, where the ventricles and corpus callosum are positioned similar to that of the MNI template. For the affine transformations (rightmost column for T1w, 2nd and 3rd rightmost columns for FA/MD maps), we expect the brain images to be aligned with that of the MNI template. (bottom) Most common failures for the pipeline are due to registration error, where either the brain images are not aligned to the MNI template or the non-linear transformation did not appropriately warp the macrostructure to match that of the MNI template.}
\end{figure}

\end{spacing}

\end{document}